\pdfoutput=1
\documentclass[twocolumn,showpacs,superscriptaddress,prb,amsmath,amssymb,floatfix,10pt,aps,eqsecnum]{revtex4-1}
\usepackage{amsmath,amssymb,color}
\usepackage{bm}
\usepackage{graphicx}
\usepackage{psfrag}
\newcommand{\bd}{\bm}

\begin{document}

\title{Singular spin-wave theory and scattering continua in the
cone state of Cs$_2$CuCl$_4$}

\author{Andreas Kreisel}
\affiliation{Department of Physics, University of Florida, Gainesville, 
Florida 32611, USA}
\affiliation{Niels Bohr Institute, 
University of Copenhagen, DK-2100 Copenhagen, Denmark}

\author{Michael Peter}
\affiliation{Institut f\"{u}r Theoretische Physik, Universit\"{a}t Frankfurt,  Max-von-Laue Strasse 1, 60438 Frankfurt, Germany}

\author{Peter Kopietz}
\affiliation{Department of Physics, University of Florida, Gainesville, 
 Florida 32611, USA}
\affiliation{Institut f\"{u}r Theoretische Physik, Universit\"{a}t
  Frankfurt,  Max-von-Laue Strasse 1, 60438 Frankfurt, Germany}

\date{July 25, 2014}

 \begin{abstract}

For temperatures below $ 0.6\,\text{K}$ the 
geometrically frustrated layered quantum antiferromagnet Cs$_2$CuCl$_4$ 
in  a magnetic field perpendicular to the layers
orders magnetically in a so-called cone state where
the magnetic moments have a finite component in the field direction, whereas
their projection onto the layers forms a spiral.
Modeling this system by a two-dimensional spatially anisotropic
quantum Heisenberg antiferromagnet with Dzyaloshinskii-Moriya interaction,
we find that even for vanishing temperatures the usual spin-wave expansion
is plagued by infrared divergencies which
are due to the coupling between longitudinal and
transverse spin fluctuations in the cone state.
Similar divergencies appear also in the
ground state of the interacting Bose gas in two and three dimensions.
Using known results for the correlation functions of the interacting Bose gas,
we present a nonperturbative  expression for the
dynamic structure factor in the cone state of  Cs$_2$CuCl$_4$. 
We show that in this state  the spectral line shape of spin fluctuations 
exhibits  singular scattering continua which
can be understood in terms of the well-known
anomalous longitudinal fluctuations in the ground state of the two-dimensional Bose gas.

\end{abstract}
\pacs{75.10.Jm, 
05.30.Jp, 
03.75.Kk, 
75.40.Gb 
}

\maketitle

\section{Introduction}

\begin{figure}
\includegraphics[width=\linewidth]{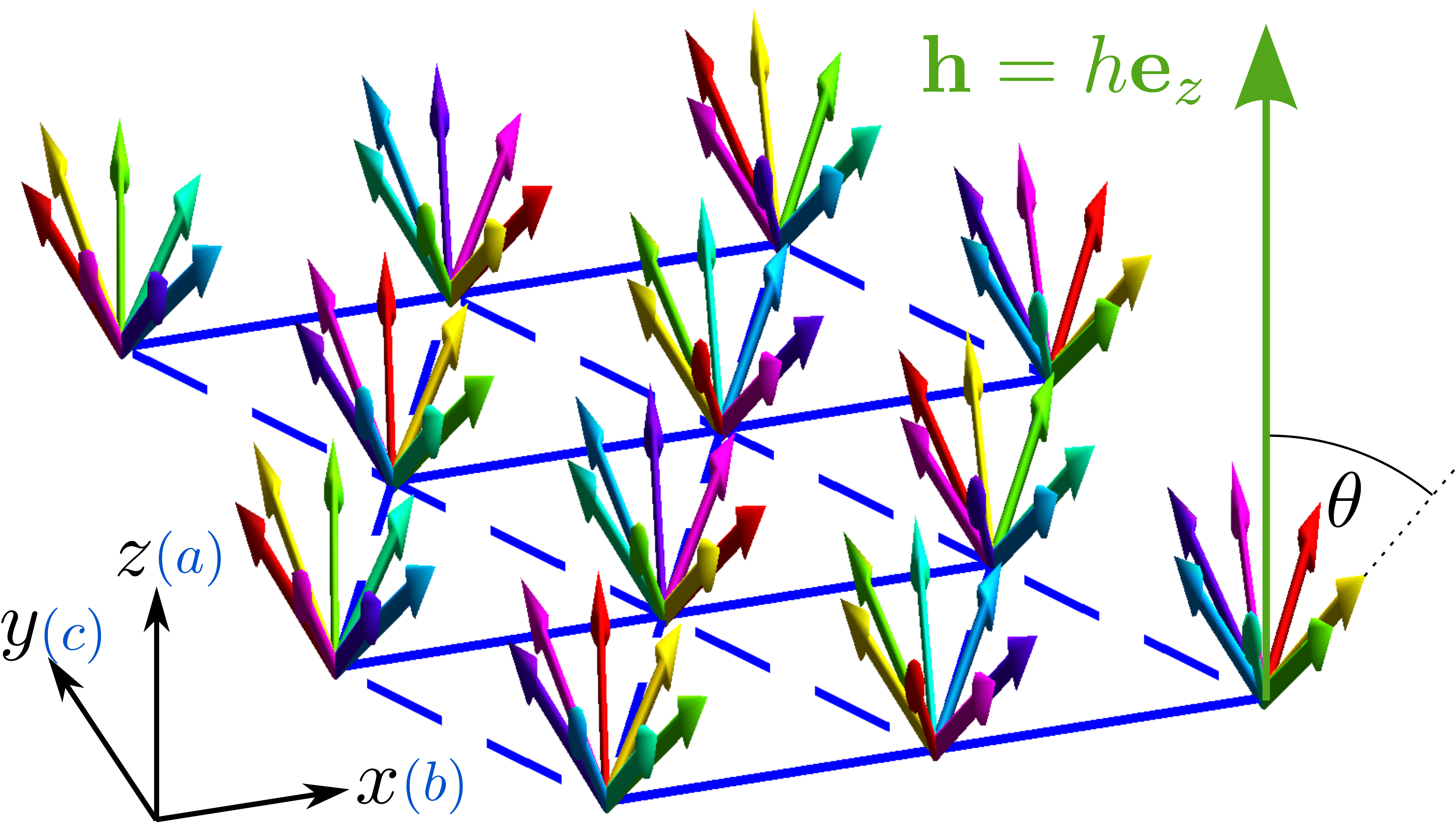}
\caption{
(Color online) 
Graphical representation of the spin configuration in the
cone state, which is the ground state of 
Cs$_2$CuCl$_4$ in a magnetic field $\bd{h}=h \hat{\bd{z}}$  pointing
along the crystallographic $a$ axis.
An explicit expression for the spin configuration in the cone state is
given in Eq.~(\ref{eq:spiral}). In the figure, 
a set of arrows with the same color
represents the cone state with incommensurate ordering vector $\bd Q$
where the spins are tilted towards the magnetic field ($\theta / \pi =0.35$ in the figure).
Different colors show the other possible states for the same value of
the magnetic field that differ only by a gauge transformation of the magnetic order
parameter.
}
\label{fig:cone}
\end{figure}

The magnetic insulator Cs$_2$CuCl$_4$ is one of the few known
physical realizations of a spin system where the combined effect of
geometric frustration and strong spin fluctuations 
can stabilize the spin-liquid phase in a substantial range of
temperatures $T$ and external magnetic fields $H$, see Ref.~[\onlinecite{Balents10}]
for a recent review.
The phase diagram of  Cs$_2$CuCl$_4$ as a function of $T$ and $H$ 
has been thoroughly mapped out using various experimental 
techniques\cite{Coldea02,Coldea03,Sytcheva09,Kreisel11,Vachon11} and 
comprises paramagnetic,  spin-liquid, and  magnetically ordered phases. 
The spin liquid phase,
which for vanishing magnetic field
is observed in the temperature  range $0.6$ K $< T <  2.6 $ K, 
has received a lot of attention from 
theory.\cite{Starykh07,Kohno07,Balents10,Starykh10,Griset11,Chen13,Herfurth13}
On the other hand, it seems to be generally accepted that the magnetically ordered 
low-temperature phase 
of Cs$_2$CuCl$_4$  in an external magnetic field along the crystallographic $a$ axis
is rather conventional.
For this direction of the magnetic field,
the magnetic moments in the ordered phase form a  so-called cone state 
(also called umbrella state), 
where the moments have a finite projection onto
the direction of the field, whereas their projection onto the plane perpendicular to the
field forms a spiral, as shown in Fig.~\ref{fig:cone}.
We will show in this work that in this phase the usual spin-wave expansion
is plagued by infrared divergencies.
 At the first sight, this is rather surprising, because
for a two-dimensional spin model
the usual phase space arguments (which are employed 
in the proof of the Mermin-Wagner theorem\cite{Mermin66}  to rule out the
existence of long-range magnetic order at finite $T$)
do not imply any divergencies at $T=0$.
We show that the infrared divergencies encountered in the cone state of
Cs$_2$CuCl$_4$ have a different physical origin: They arise from the
coupling between longitudinal and gapless transverse spin fluctuations.
In fact, similar divergences are encountered in the theory of the interacting Bose gas
where interaction corrections to
Bogoliubov's celebrated mean-field
theory for the condensed phase are known to be infrared divergent
in two and three dimensions.\cite{Nepomnyashchii78,Popov79,Weichman88,Giorgini92,Castellani97,Pistolesi04}
Note, however, that these divergencies do not arise in the spin-wave
expansion of many other quantum magnets because spin conservation
forces this coupling to vanish in the long wavelength limit.
Fortunately, in  the Bose gas nonperturbative resummations of these divergencies
are available, leading to a nonanalytic contribution to the longitudinal 
part of the bosonic two-point function.\cite{Sachdev99b,Zwerger04,Kreisel07,Dupuis11}
In this work we will
apply
these results to the cone state of Cs$_2$CuCl$_4$  and show that they imply
the existence of
extended scattering continua on the spin dynamic structure factor 
in a magnetic field.  A similar strategy has been adopted previously
in Ref.~[\onlinecite{Kreisel07}] to calculate the
dynamic structure factor of a quantum antiferromagnet in a uniform magnetic field.

\begin{figure}
 \includegraphics[width=\linewidth]{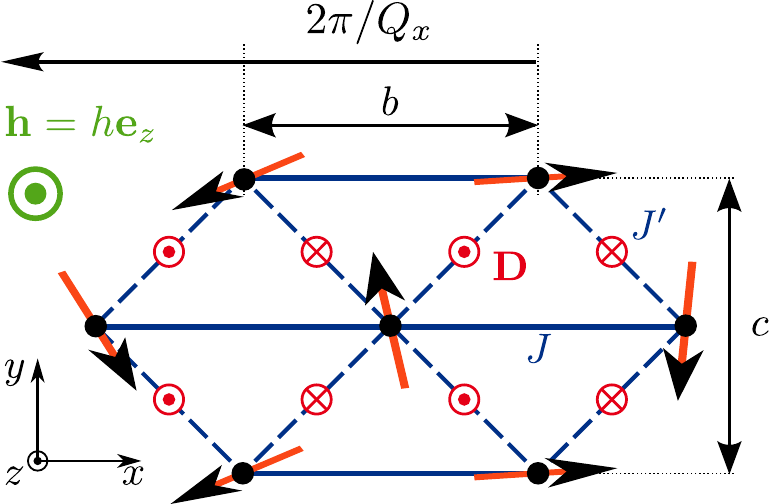}
\caption{(Color online) 
Projection of the spin configuration in the cone state on the
plane perpendicular to the magnetic field  (the $x$-$y$ plane).
The arrows represent the magnetic moments.
The spiral state is characterized by the ordering vector 
$\bd Q=Q_x\hat{\bd{x}}$. Apart from the antiferromagnetic exchange interactions 
$J$ and $J'$ there are also Dzyaloshinskii-Moriya interactions 
$\bd D=\pm D \hat{\bd{z}}$.
Small circles with dots (crosses) denote vectors pointing out of (into) the
 $x$-$y$ plane.
}
\label{fig_model}
\end{figure}

The minimal model for describing the cone state
in Cs$_2$CuCl$_4$ is the following two-dimensional
antiferromagnetic Heisenberg model in an external magnetic field
along the crystallographic $a$ axis (which we identify with the $z$ axis
of our coordinate system)
\begin{align}
 {\cal{H}}   & =  \frac{1}{2} \sum_{ i j  } 
 \left[ 
J_{ij} {\bd{S}}_i \cdot {\bd{S}}_j +
 \bd{D}_{ ij} \cdot (  {\bd{S}}_i \times {\bd{S}}_j ) \right]
  -      h \sum_{i} {{S}}^z_i.
 \label{eq:Hs}
 \end{align}
Here the sums are over all $N$ sites of the lattice, and
$h = g \mu_B H$ is the Zeeman energy associated with the magnetic field $H$,
where $g \approx 2.19$ is the effective Land\'e factor\cite{Coldea02} and $\mu_B$ is the Bohr 
magneton. The spin operators 
$\bd{S}_i = \bd{S} ( \bd{R}_i )$ satisfy $\bd{S}_i^2 = S( S+1)$ with $S=1/2$ and
are localized at the sites $\bd{R}_i$
of a distorted triangular Bravais lattice
with crystallographic lattice constants $b$ and $c$
as shown in Fig.~\ref{fig_model}.
The exchange couplings $J_{ij} = J ( \bd{R}_i - \bd{R}_j )$ connect nearest neighbors 
on the distorted triangular lattice with $ J ( \pm \bd{\delta}_1 ) = J =0.374\,\text{meV} = 
4.34 \, \text{K} $ and
$ J ( \pm \bd{\delta}_2 ) =   J ( \pm \bd{\delta}_3 )  = J^{\prime}  =0.128\,\text{meV}   
= 1.49 \, \text{K} $
where the three elementary direction vectors are
 \begin{align}
 \bd{\delta}_1 & =   b \hat{\bd{x}},
 \; \; \; 
 \bd{\delta}_2  =   - \frac{b}{2} \hat{\bd{x}} + \frac{c}{2} \hat{\bd{y}},
 \; \; \;
  \bd{\delta}_3 =   - \frac{b}{2} \hat{\bd{x}} - \frac{c}{2} \hat{\bd{y}}.
 \end{align}
Here $\hat{\bd{x}}$, $\hat{\bd{y}}$ and $\hat{\bd{z}}$ are
unit vectors around the three Cartesian directions of our coordinate system. 
Due to the low symmetry of the crystal the spins are additionally coupled by an
antisymmetric Dzyaloshinskii-Moriya interaction 
of the form $\bd{D}_{ij} = D_{ij} \hat{\bd{z}} $, where
$ D_{ij} = D ( \bd{R}_i - \bd{R}_j )$ is finite if $\bd{R}_i$ and $\bd{R}_j$
are nearest neighbors along the diagonal bonds.\cite{Starykh10} The numerical value
of the nearest-neighbor Dzyaloshinskii-Moriya coupling is
$D ( \pm \bd{\delta}_2 ) = D ( \pm \bd{\delta}_3 ) = \mp D $ with $D = 
0.020\,\text{meV} = 0.23 \, \text{K}$.

The above spin model has  continuous rotational symmetry with respect to
rotations around the $z$ axis in spin space. As a consequence, one of the
magnons is gapless, and the system does not exhibit any long-range order
at any finite temperature.\cite{Mermin66}  If one nevertheless tries to
calculate the magnetization at finite $T$ using spin-wave theory, one
encounters an infrared divergence signaling the
inconsistency of the assumption of finite magnetization.
On the other hand, at vanishing temperature one should not expect any
infrared divergencies because the fluctuations are not strong enough
to destroy long-range magnetic order.
In this work we will show that the spin-wave expansion for this model is nevertheless plagued by
infrared divergencies which have to be resummed to all orders to obtain
meaningful results. 
Similar singularities in the spin-wave expansion of the magnetic anisotropy energy
of quantum antiferromagnets
have been discussed  some time ago by Maleyev\cite{Maleyev00} and by
Syromyatnikov and Maleyev;\cite{Syromyatnikov01} however, these authors did not
offer any strategy to resum these divergencies in order to
obtain physical results.

The rest of this paper is organized as follows:
In Sec.~\ref{sec:spinwave} we describe the spin-wave expansion in the cone state of
Cs$_2$CuCl$_4$. To identify and isolate the infrared divergencies,
we use a special parametrization of the spin-wave expansion
in terms of Hermitian operators,\cite{Hasselmann06,Kreisel07,Kreisel08,Kreisel11} which
explicitly separates the longitudinal from the transverse spin fluctuations.
In Sec.~\ref{sec:bose} we briefly recall some relevant results for the correlation functions
of the interacting Bose gas and identify the corresponding parameters for the
cone state of Cs$_2$CuCl$_4$. 
In Sec.~\ref{sec:spinstruc} we use the mapping of the bosonized 
spin-wave Hamiltonian in the cone state
onto the interacting Bose gas to discuss the
dynamic structure factor of   Cs$_2$CuCl$_4$, which can be directly measured
by means of neutron scattering. In particular, we show that
in a magnetic field the dynamic structure factor 
exhibits scattering continua
which are directly related to the well-known anomalous longitudinal fluctuations 
of the interacting Bose gas.
Finally, in  Sec.~\ref{sec:conclusions} we summarize our results
and discuss ways to verify our theoretical prediction experimentally.

\section{Divergent spin-wave expansion in the cone state}
\label{sec:spinwave}

In this section we shall set up the spin-wave expansion in the cone state of
Cs$_2$CuCl$_4$ and identify the infrared divergent contributions. 
We first follow the conventional formulation\cite{Kreisel11} in terms of canonical
boson operators introduced via the Holstein-Primakoff 
transformation\cite{Holstein40} 
and then show that an alternative parametrization
of the $1/S$ expansion using
Hermitian operators\cite{Hasselmann06,Kreisel07,Kreisel08,Kreisel11} 
allows us to isolate the
infrared-divergent terms in a very efficient way.

\subsection{Expansion in terms of canonical bosons}

Before we set up the spin-wave expansion, we should determine the
spin configuration in the classical ground state.
Therefore we replace the spin operators by classical vectors of length $S=1/2$ 
pointing in the direction of the local magnetization.\cite{Veillette05,Kreisel11}
For the triangular lattice
antiferromagnet it has been shown that a finite
magnetization in a spiral state is stable even in the presence of quantum fluctuations
if the exchange parameters are in the range\cite{Trumper99,Note1}
$0.27 J \lesssim J' < 2J$.
Taking into account the Dzyaloshinskii-Moriya anisotropy with $D>0$, the classical ground-state
energy,
\begin{equation}
  {\cal{H}}_0  = 
  \frac{S^2}{2} \sum_{  ij}   J^{\parallel} _{ij}  
  -  S \sum_{i  }  {\bd{h}}  \cdot \hat{\bd{m}}_i \,,
  \label{eq:H0} 
\end{equation}
is lowered compared to the situation without anisotropy, thus stabilizing
the spiral state. Here we have defined
 \begin{equation}
J_{ij}^{\parallel}  =  J_{ij}  
  \hat{\bd{m}}_i \cdot \hat{\bd{m}}_j +
 {\bd{D}}_{ij} \cdot (  \hat{\bd{m}}_i \times \hat{\bd{m}}_j ).
 \end{equation}
In the presence of a magnetic field, the classical ground state is
the so-called cone state, where the spins form
a spiral that is tilted towards the direction of the field such that the magnetization $\hat{\bd{m}}_i$  on lattice point $ {\bd{R}}_i$ is given by\cite{Veillette05,Kreisel11}
\begin{equation}
 \hat{\bd{m}}_i  = 
 \sin \theta      [ \cos ( \bd{Q} \cdot \bd{R}_i )    \hat{\bd{x}} +  
   \sin ( \bd{Q} \cdot \bd{R}_i )  \hat{\bd{y}} ] +
    \cos \theta     \hat{\bd{z}} .
 \label{eq:spiral}
 \end{equation}
This cone state is characterized by the opening angle $2 \theta$
of the cone (see Fig.~\ref{fig:cone}) and the wave vector $\bd{Q}$ of the spiral.
Note that in Ref.~[\onlinecite{Kreisel11}] a different definition of the angle $\theta$
has been used, which amounts to the re-definition $\theta \rightarrow \pi - \theta$.
The parameters $\theta$ and $\bd{Q}$ should be determined by minimizing
classical ground state energy in Eq. (\ref{eq:H0}).
Anticipating that the spiral wave vector is of the form\cite{Note2}
$\bd{Q}=Q_x \hat{\bd{x}}$ as indicated in Fig.~\ref{fig_model},
it is easy to show that
the classical ground-state energy is minimal if the angle $\theta$ is given by
 \begin{equation}
\cos \theta = h/ h_c,
 \label{eq:tiltclassical}
 \end{equation}
whereas the wave vector of the spiral obeys the equation,
 \begin{equation}
\cos \left( \frac{Q_x b }{2} \right) = - \frac{J^{\prime}}{2J} -
 \frac{D}{2 J} \cot \left( \frac{Q_x b}{2} \right)\,.
 \end{equation}
Here the critical magnetic field is given by
\begin{equation}
 h_c = S ( J^{D}_0 - J^{D}_{\bd{Q}}) = S( J_0 - J_{\bd{Q}} + i D_{\bd{Q}}),
 \label{eq:hcdef} 
\end{equation}
where we have introduced the Fourier transforms of the
exchange integrals and the Dzyaloshinskii-Moriya interaction,
\begin{align}
 J_{\bd{k}}  =&
 2 J \cos ( k_x b ) + 4 J^{\prime} \cos \left( \frac{k_x b}{2} \right) 
\cos \left(
 \frac{  k_y c}{2} \right) ,
 \\
 {{D}}_{\bd{k}} 
= & - 4 i D  \sin \left( \frac{k_x b}{2} \right) 
\cos \left(
 \frac{  k_y c}{2} \right).
  \end{align}
For later convenience, we have also introduced the
real quantity,
  \begin{equation}
  J_{\bd{k}}^D = J_{\bd{k}} - i D_{\bd{k}}.
 \label{eq:FQdef}
 \end{equation}

To set up the $1/S$ expansion, we project the spin operators
at each lattice site $\bd{R}_i$  onto a local  basis whose $z$ axis matches the direction of the
classical ground state,
\begin{equation}
 \bd{S}_i=S_i^{\parallel}\bd{\hat{m}}_i+S_i^{(1)}\bd{e}_i^{(1)}+S_i^{(2)}\bd{e}_i^{(2)}.
 \label{eq:Sexpansion}
\end{equation}
The convenient choice for the transverse basis vectors is\cite{Kreisel11}
\begin{align}
 \bd{e}^{(1)}_i
 & =    \sin ( \bd{Q} \cdot \bd{R}_i )    \hat{\bd{x}}   
   - \cos ( \bd{Q} \cdot \bd{R}_i )  \hat{\bd{y}},
 \label{eq:ei1}
 \\
  \bd{e}^{(2)}_i & = 
\cos \theta   [ \cos ( \bd{Q} \cdot \bd{R}_i )    \hat{\bd{x}} +  
   \sin ( \bd{Q} \cdot \bd{R}_i )  \hat{\bd{y}} ] -
  \sin \theta  \; \hat{\bd{z}} .
 \label{eq:ei2}
 \end{align}
We then evaluate all scalar products and cross products of the basis vectors 
and express the components of the spin operators in terms of canonical
boson operators $b_i$ and $b_i^{\dagger}$ using the Holstein-Primakoff
transformation,\cite{Holstein40}
 \begin{subequations}
 \begin{align}
 S_i^{\parallel} & =  S- n_i,
 \\
 S_i^+ &=  \sqrt{2S} \sqrt{ 1 - \frac{ n_i}{2S}} b_i ,
 \\
 S_i^- &=  \sqrt{2S} b^{\dagger}_i \sqrt{ 1 - \frac{ n_i}{2S}}  ,
\end{align}
\end{subequations}
where $n_i = b^{\dagger}_i b_i$.
An expansion of the square roots generates the $1/S$ expansion and allows us
to rewrite the Hamiltonian as
\begin{equation}
 {\cal{H}} = {\cal{H}}_0 + \sum_{n=2}^{\infty} {\cal{H}}_n\,,
\label{eq:hexpansion}
\end{equation}
with ${\cal{H}}_n\propto S^{2-n/2}$.
The quadratic term describes noninteracting spin waves; in terms of the
Fourier transforms of the boson operators,
 \begin{equation}
 b_{\bd{k}} = \frac{1}{\sqrt{N}} \sum_{i} e^{ -i \bd{k} \cdot {\bd{R}}_i } b_{i},
 \end{equation}
we obtain
\begin{align}
 {\cal{H}}_2 & =  \sum_{\bd{k}}  \Bigl\{  A_{\bd{k}}
b^{\dagger}_{\bd{k}} b_{\bd{k}} 
 - \frac{ B_{\bd{k}}}{2} \bigl[ 
 b^{\dagger}_{\bd{k}} b^{\dagger}_{-\bd{k}} +
b_{-\bd{k}} b_{\bd{k}} \bigr] \Bigr\},
 \label{eq:H2bb}
 \end{align} 
where $A_{\bd{k}} = A_{\bd{k}}^{+} + A_{\bd{k}}^{-}$ and
 \begin{subequations}
 \begin{align}
 A^{+}_{\bd{k}} & =     B_{\bd{k}}
- S
 \left[  J^{D}_{\bd{Q}} - \frac{ J^{D}_{ \bd{Q} + \bd{k}} +   J^{D}_{ \bd{Q} - \bd{k}} }{2} 
\right] =A^+_{-\bd{k}}, 
 \label{eq:Akdef}
 \\
 A^-_{\bd{k}} & =      S  \cos {\theta}
\frac{  J^{D}_{ \bd{Q} + \bd{k}} -   J^{D}_{ \bd{Q} - \bd{k}} }{2} =-A^-_{-\bd{k}},
 \label{eq:Akminusdef}
\\
  B_{\bd{k}} & =       
\frac{S}{2}     \sin^2 \theta
 \left[  J_{\bd{k}} - \frac{ J^{D}_{ \bd{Q} + \bd{k}} +   J^{D}_{ \bd{Q} - \bd{k}} }{2} \right]
 =B_{-\bd{k}}.
 \label{eq:Bkdef}
\end{align}
 \end{subequations}
The cubic term ${\cal{H}}_3$ describes the leading
spin-wave interaction processes. In momentum space it can be written as\cite{Kreisel11}
 \begin{eqnarray}
{\cal{H}}_3  & = &  \frac{1}{\sqrt{N} }
 \sum_{ \bd{k}_1 \bd{k}_2 \bd{k}_3} \delta_{ \bd{k}_1 +
 \bd{k}_2 + \bd{k}_3  , 0 } \Bigl[
 \nonumber
 \\
 &  & \frac{1}{ 2!}
\Gamma_3^{ {b}^{\dagger} {b}^{\dagger} b }
 ( \bd{k}_1 , \bd{k}_2 ; \bd{k}_3 )
 b^{\dagger}_{  - \bd{k}_1 } b^{\dagger}_{  - \bd{k}_2 }
 b_{  \bd{k}_3 }  
 \nonumber
 \\
 & + & \frac{1}{ 2!} 
  \Gamma_3^{ {b}^{\dagger} bb }
 ( \bd{k}_1 ; \bd{k}_2 , \bd{k}_3 )
 b^{\dagger}_{ - \bd{k}_1 } b_{  \bd{k}_2 }
 b_{ \bd{k}_3 }
\Bigr].
 \label{eq:H3hp}
 \end{eqnarray}
The properly symmetrized cubic interaction
vertices are
 \begin{subequations}
 \begin{align}
\Gamma_3^{ {b}^{\dagger} {b}^{\dagger} b }
 ( \bd{k}_1 , \bd{k}_2 ; \bd{k}_3 ) & = 
   -  \sin {\theta}  \frac{ \sqrt{2S}}{2 i }
 \left[ K_{-\bd{k}_1} + K_{ - \bd{k}_2 } \right], 
 \label{eq:Gammabbb1}
\\
 \Gamma_3^{ {b}^{\dagger} b b }
 ( \bd{k}_1 ; \bd{k}_2 , \bd{k}_3 ) & = 
   \sin {\theta}  \frac{ \sqrt{2S}}{2 i }
 \left[ K_{\bd{k}_2} + K_{  \bd{k}_3 } \right],
 \label{eq:Gammabbb2}
\end{align}
 \end{subequations}
where
 \begin{equation}
 K_{\bd{k}} = \cos \theta 
\left[  J_{\bd{k}} - \frac{ J^{D}_{ \bd{Q} + \bd{k}} +   J^{D}_{ \bd{Q} - \bd{k}} }{2} \right] -
\frac{  J^{D}_{ \bd{Q} + \bd{k}} -   J^{D}_{ \bd{Q} - \bd{k}} }{2}.
 \label{eq:Kkdef}
 \end{equation}
Finally, the quartic part ${\cal{H}}_4$ of the interaction can be written as
\begin{align}
 {\cal{H}}_4  & =  \frac{1}{N }
 \sum_{ \bd{k}_1 \bd{k}_2 \bd{k}_3 \bd{k}_4 }
 \delta_{ \bd{k}_1 +
 \bd{k}_2 + \bd{k}_3 + \bd{k}_4 , 0 } \Bigl[
 \nonumber
 \\
 &  \frac{1}{(2!)^2}
\Gamma_4^{ {b}^{\dagger} {b}^{\dagger} b b}
 ( \bd{k}_1 , \bd{k}_2 ; \bd{k}_3 , \bd{k}_4)
 b^{\dagger}_{  - \bd{k}_1 } b^{\dagger}_{  - \bd{k}_2 }
 b_{  \bd{k}_3} b_{  \bd{k}_4}  
 \nonumber
 \\
 & +  \frac{1}{ 3!} 
  \Gamma_4^{ {b}^{\dagger} bbb }
 ( \bd{k}_1 ; \bd{k}_2 , \bd{k}_3 ,\bd{k}_4 )
 b^{\dagger}_{ - \bd{k}_1 } b_{  \bd{k}_2 }
 b_{ \bd{k}_3} b_{ \bd{k}_4 }\nonumber
 \\
& +  \frac{1}{ 3!} 
  [\Gamma_4^{ {b}^{\dagger} bbb}
 ( \bd{k}_1 ; \bd{k}_2 , \bd{k}_3 ,\bd{k}_4 )]^*
 b^{\dagger}_{ - \bd{k}_4 } b^{\dagger}_{ - \bd{k}_3 }
 b^{\dagger}_{ -\bd{k}_2} b_{ \bd{k}_1 }
\Bigr].
 \label{eq:H4hp}
 \end{align}
The  vertices are
 \begin{subequations}
 \begin{align}
\Gamma_4^{ {b}^{\dagger} {b}^{\dagger} bb }
 ( \bd{k}_1 , \bd{k}_2 ; \bd{k}_3 , \bd{k}_4 ) & =
-\frac 18 \sum_{i=1}^4(J_{\bd{k}_i}^{\pm}+J_{-\bd{k}_i}^\mp)\notag\\
&\hspace{-25mm}+\frac 12 (J_{\bd{k}_1-\bd{k}_3}^\parallel+J_{\bd{k}_2-\bd{k}_3}^\parallel+
J_{\bd{k}_1-\bd{k}_4}^\parallel+J_{\bd{k}_2-\bd{k}_4}^\parallel), 
 \label{eq:Gammabbbb1}
\\
 \Gamma_4^{ {b}^{\dagger} b b b}
 ( \bd{k}_1 ; \bd{k}_2 , \bd{k}_3,  \bd{k}_4) & =-\frac 14(J_{\bd{k}_2}^{++}+
J_{\bd{k}_3}^{++}+J_{\bd{k}_4}^{++}),
 \label{eq:Gammabbbb2}
\end{align}
\end{subequations}
with
 \begin{subequations}
\begin{eqnarray}
 J^{\parallel}_{\bd{k}} & = & 
\cos^2\theta J_{\bd{k}} + \sin^2 {\theta}
 \frac{ J^{D}_{ \bd{Q} + \bd{k}} + J^{D}_{\bd{Q} - \bd{k}} }{2} ,
 \label{eq:JparallelFT}
 \\
J^{+-}_{\bd{k}} &=& J^{-+}_{- \bd{k}}    = 
\sin^2{\theta} J_{\bd{k}} + (1+ \cos^2 {\theta})
 \frac{ J^{D}_{ \bd{Q} + \bd{k}} + J^{D}_{\bd{Q} - \bd{k}} }{2} 
 \nonumber
 \\
 & &\hspace{1.3cm} + \cos {\theta} \left[ J^{D}_{ \bd{Q} + \bd{k}} - J^{D}_{\bd{Q} - \bd{k}} \right],
 \\
J^{++}_{\bd{k}} &=& J^{--}_{\bd{k}}    =  - \sin^2 {\theta} \left[
J_{\bd{k}} -
 \frac{ J^{D}_{ \bd{Q} + \bd{k}} + J^{D}_{\bd{Q} - \bd{k}} }{2} \right] .
 \label{eq:JtransverseFT}
 \end{eqnarray}
\end{subequations}

\begin{figure}
\includegraphics[width=\linewidth]{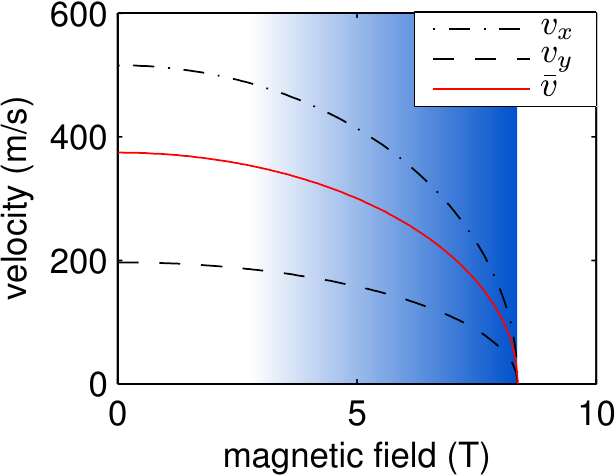}
\caption{(Color online) 
Plot of the magnon velocities $v_x$ (dashed line) and $v_y$ (dashed-dotted line)
defined in Eqs.~(\ref{eq:vx}) and(\ref{eq:vy}) for the parameters relevant
for Cs$_2$CuCl$_4$. The solid (red) line represents the angular averaged velocity defined
in Eq.~(\ref{eq:angular}).
The shaded (blue) area is the regime close to the saturation field where our mapping to the Bose gas
is quantitatively accurate.
}
\label{fig:velocities}
\end{figure}

To obtain  the magnon dispersion of our model,
we diagonalize ${\cal{H}}_2$ in Eq.~(\ref{eq:H2bb}) by means of a 
Bogoliubov transformation and obtain\cite{Kreisel11}
\begin{equation}
 {\cal{H}}_2 = \sum_{\bd{k}} \left[ E_{\bd{k}} \beta^{\dagger}_{\bd{k}} \beta_{\bd{k}} +
 \frac{ \epsilon_{\bd{k}} - A^+_{\bd{k}}}{2} \right]\,,
 \label{eq:H2diag}
 \end{equation}
where $\beta^{\dagger}_{\bd{k}}$ and $\beta_{\bd{k}}$ are again canonical boson 
operators and the  magnon dispersion is
 \begin{equation}
 E_{\bd{k}} =
 \epsilon_{\bd{k}} + A^-_{\bd{k}} ,
 \label{eq:magdispersion}
 \end{equation}
with
 \begin{equation}
 \epsilon_{\bd{k}}
= \sqrt{ (A^+_{\bd{k}})^2 -  B_{\bd{k}}^2 }\,.
 \label{eq:epsdispersion}
 \end{equation}
Using the fact that for small wave vectors the antisymmetric contribution is negligible,
\begin{equation}
 A^{-}_{\bd{k}}
=  {\mathcal{O}} ( \bd{k}^3 ) ,
 \label{eq:Aminussmall}
 \end{equation}
it is easy to show that at long wavelengths the magnons have a linear dispersion,
\begin{equation}
 E_{\bd{k}} =  \epsilon_{\bd{k}} + {\mathcal{O}} ( \bd{k}^3 ) = v ( {\hat{\bd{k}}}  ) | \bd{k} |  + {\mathcal{O}} ( \bd{k}^3 ) ,
 \label{eq:Eksmall}
 \end{equation}
with direction-dependent magnon velocity
\begin{equation}
v ( {\hat{\bd{k}}}  ) = \sqrt{ v_x^2 \hat{k}_x^2 + v_y^2 \hat{k}_y^2 },
 \label{eq:vkhat}
 \end{equation}
where $\hat{\bd{k}} = \bd{k} / | \bd{k} |$.
The Cartesian components of the squares of the magnon velocity are explicitly
  \begin{align}
 v_x^2 & =   S h_c b^2 \sin^2 \theta
 \biggl[ - J \cos ( Q_x b ) - \frac{ J^{\prime}}{2}  \cos \Bigl( \frac{Q_x b}{2} \Bigr) 
 \nonumber
 \\
 & 
 \hspace{20mm} + \frac{D}{2} 
  \cos \Bigl( \frac{Q_x b}{2}  \Bigr) \biggr],
 \label{eq:vx}
 \\
  v_y^2 & =   S h_c c^2 \sin^2 \theta
 \biggl[ - \frac{ J^{\prime}}{2}  \cos \Bigl( \frac{Q_x b}{2} \Bigr) 
+ \frac{D}{2} 
  \cos \Bigl( \frac{Q_x b}{2}  \Bigr) \biggr].
 \nonumber
 \\
 & & 
 \label{eq:vy}
 \end{align}
A graph of the velocity components $v_x$ and $v_y$ 
as a function of the magnetic field is shown in  Fig.~\ref{fig:velocities}.
An important observation is that the cubic interaction vertices in 
Eqs.~(\ref{eq:Gammabbb1}) and (\ref{eq:Gammabbb2}) have 
finite limits when all wave-vectors vanish. Using the fact that
 \begin{equation}
 K_0 =  \cos \theta  ( J_0 - J_{\bd{Q}}^D) =  \cos {\theta} \frac{h_c}{S} = \frac{h}{S},
 \label{eq:K0lim}
 \end{equation}
we obtain
 \begin{align}
\Gamma_3^{ {b}^{\dagger} {b}^{\dagger} b }
 ( 0 , 0 ; 0 ) & = - \Gamma_3^{ {b}^{\dagger} b b }
 ( 0 ; 0 , 0 ) 
 \nonumber
 \\
 & =
     2 i  \sin \theta \cos {\theta}  \frac{ h_c}{  \sqrt{2S}}  .
 \label{eq:Gammabbb1k}
\end{align}
In the next section we will show that
the linear magnon dispersion in combination 
with finite cubic interaction vertices imply that the
magnon self-energy generated by the spin-wave interactions is infrared divergent.

\subsection{Expansion in terms of Hermitian field operators}
\label{subsec:hermit}

To calculate corrections to linear spin-wave theory in powers 
of the formal small parameter $1/S$,
we now use conventional many-body methods for bosons.
We find that the perturbation series for the magnon self-energy
contains already at order $1/S$ some infrared-divergent terms. 
Unfortunately, in the conventional formulation of the spin-wave expansion
using the Bogoliubov bosons $\beta_{\bd{k}}$ and $\beta_{\bd{k}}^{\dagger}$,
which diagonalize the quadratic Hamiltonian
 the calculations become rather cumbersome
because the interactions in this basis
contain the singular coefficients of the Bogoliubov transformation
which is necessary to bring the quadratic Hamiltonian (\ref{eq:H2bb}) to the
diagonal form (\ref{eq:H2diag}).
One can avoid this problem by directly 
generating the perturbation expansion in terms of the
Holstein-Primakoff bosons,\cite{Maleyev00}
but then one has to deal with anomalous propagators and off-diagonal magnon self-energies.
Fortunately, there is an alternative parametrization of the spin-wave expansion
using Hermitian operators\cite{Hasselmann06, Kreisel07, Kreisel08, Kreisel11}
which greatly facilitates the identification of the singular
terms in the spin-wave expansion around the cone state of  Cs$_2$CuCl$_4$.
In this approach, one expresses the Holstein-Primakoff bosons 
in terms of Hermitian operators $\Pi_{\bd k}$ and $\Phi_{\bd k}$ by setting
\begin{equation}
 b_{\bd k}=\frac 1{\sqrt 2} [\Phi_{\bd k}+i\Pi_{\bd k}]\;,
 \; \; \; 
 b_{\bd k}^{\dagger} =\frac 1{\sqrt 2} [ \Phi_{-\bd k} - i \Pi_{-\bd k}]\;.
 \label{eq:hermit}
\end{equation}
By demanding that
$ [\Phi_{\bd k} ,\Pi_{\bd k}]=i\delta_{\bd k, -\bd k'}$ and that all other commutators vanish,
it is easy to see that the canonical commutation relations $[ b_{\bd{k}} , b^{\dagger}_{\bd{k}^{\prime}} ] = \delta_{\bd{k} , \bd{k}^{\prime}}$ are satisfied.
Then we obtain from Eq.~(\ref{eq:H2bb}) for the 
quadratic part of the spin-wave Hamiltonian,
 \begin{eqnarray}
 {\cal{H}}_2  & = & \frac{1}{2} \sum_{\bd{k}} \Bigl[
 \Delta_{\bd{k}} {\Pi}_{- \bd{k} } {\Pi}_{\bd{k}} 
 + \frac{ \epsilon_{\bd{k}}^2}{ \Delta_{\bd{k}}}
  {\Phi}_{ - \bd{k} } {\Phi}_{\bd{k}} - A^+_{\bd{k}}
 \nonumber
 \\
 &  & \hspace{7mm} + i A^-_{\bd{k}} 
  ( {\Phi}_{ - \bd{k} }  {\Pi}_{\bd{k}} - 
  {\Pi}_{ - \bd{k} } {\Phi}_{ \bd{k} } ) \Bigr],
 \label{eq:H2swphi}
 \end{eqnarray}
where $\epsilon_{\bd{k}}$ and $A_{\bd{k}}^-$ are  defined in Eq.~(\ref{eq:epsdispersion}) and
(\ref{eq:Akminusdef}), and
 \begin{equation}
 \Delta_{\bd{k}} = A_{\bd{k}}^+ +  B_{\bd{k}}.
 \end{equation}
Using Eqs.~(\ref{eq:Akdef}) and (\ref{eq:Bkdef}) we find that
for $\bd{k} \rightarrow 0$ the energy scale $\Delta_{\bd{k}}$ approaches a finite limit,
 \begin{equation}
 \Delta_{0} = h_c \sin^2 \theta  = h_c \left[ 1 - \frac{ h^2}{h_c^2} \right],
 \end{equation}
 such that the operator $\Phi_{\bd k}$ is associated with the
transverse fluctuations and $\Pi_{\bd k}$ is associated with the longitudinal
fluctuations as it will become more clear in Sec. \ref{sec:bose}.
The cubic part ${\cal{H}}_3$ of our spin-wave Hamiltonian
in Eq.~(\ref{eq:H3hp}) can be written as
 \begin{eqnarray}
 {\cal{H}}_3 & = & \frac{1}{\sqrt{N}} 
 \sum_{ \bd{k}_1 \bd{k}_2 \bd{k}_3} \delta_{ \bd{k}_1 +
 \bd{k}_2 + \bd{k}_3  , 0 } \Bigl[
 \nonumber
 \\
 & & \frac{1}{ 3!}
\Gamma^{ \Phi \Phi \Phi }
 ( \bd{k}_1 , \bd{k}_2 , \bd{k}_3 )
  {\Phi}_{   \bd{k}_1 } {\Phi}_{   \bd{k}_2 }
 {\Phi}_{  \bd{k}_3 }  
 \nonumber
 \\
 & + & 
\frac{1}{ 3!}
\Gamma^{ \Pi \Pi \Pi }
 ( \bd{k}_1 , \bd{k}_2 , \bd{k}_3 )
  {\Pi}_{   \bd{k}_1 } {\Pi}_{   \bd{k}_2 }
 {\Pi}_{  \bd{k}_3 }  
 \nonumber
 \\
 & + & 
\frac{1}{ 2!}
\Gamma^{ \Phi \Phi \Pi }
 ( \bd{k}_1 , \bd{k}_2 ; \bd{k}_3 )
   {\Phi}_{   \bd{k}_1 } {\Phi}_{   \bd{k}_2 }
  {\Pi}_{  \bd{k}_3 }  
 \nonumber
 \\
 & + & 
\frac{1}{ 2!}
\Gamma^{ \Pi \Pi \Phi }
 ( \bd{k}_1 , \bd{k}_2 ; \bd{k}_3 )
  {\Pi}_{   \bd{k}_1 } {\Pi}_{   \bd{k}_2 }
  {\Phi}_{  \bd{k}_3 }  
\Bigr].
 \label{eq:H3hermitian}
 \end{eqnarray}
The properly symmetrized vertices are
\begin{subequations}
 \begin{align}
  \Gamma^{\Phi\Phi\Phi}(\bd k_1,\bd k_2,\bd k_3)=&-i\left[V_-(\bd k_1)+V_-(\bd k_2)+V_-(\bd k_3)\right]\,,\\
  \Gamma^{\Pi\Pi\Pi}(\bd k_1,\bd k_2,\bd k_3)=&V_+(\bd k_1)+V_+(\bd k_2)+V_+(\bd k_3),\\
  \Gamma^{\Pi\Pi\Phi}(\bd k_1,\bd k_2;\bd k_3)=&-iV_-(\bd k_3)\,,\\
  \Gamma^{\Phi\Phi\Pi}(\bd k_1, \bd k_2 ; \bd k_3)=&V_+(\bd k_3)\,,
 \label{eq:vertexphiphipi}
 \end{align}
\end{subequations}
where
\begin{equation}
 V_\pm ( \bd{k} ) = \sqrt{S} \sin \theta \frac{K_{\bd k}\pm K_{-\bd k}}{2}\,,
\end{equation}
and  $K_{\bd{k}}$ is defined in Eq.~(\ref{eq:Kkdef}).
For small wave vectors we find
 \begin{align}
 V_{+} ( \bd{k} ) & =  \sin \theta \cos \theta \frac{h_c}{\sqrt{S}}  + {\cal{O}} ( \bd{k}^2 ),
 \\
 V_{-} ( \bd{k} ) & =  {\cal{O}} ( \bd{k}^3 ).
 \end{align}
In deriving Eq.~(\ref{eq:H3hermitian}) we have ignored terms linear
in the field operators which are related to the
specific ordering of the operators in this equation.
These terms do not affect the infrared divergencies discussed below.

\begin{figure}
  \includegraphics[width=\linewidth]{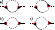}
\caption{(Color online) Graphical representation of the singular contributions to the magnon
self-energies. The diagrams (a) in the first line
give the singular contribution to $\Sigma^{\Phi \Phi} ( K )$ in Eq.~(\ref{eq:sigma1}).
Diagrams (b) and (c) represent the singular contributions to
 $\Sigma^{\Pi \Pi} ( K )$ and
 $\Sigma^{\Phi \Pi} ( K )$ given in Eqs.~(\ref{eq:sigma2}) and (\ref{eq:sigma3}).
Solid lines are propagators $G^{\Phi \Phi}$, dashed lines are
propagators $G^{\Pi \Pi}$, and the solid-dashed lines represent the  
off-diagonal propagators
$G^{\Phi \Pi}$ and $G^{\Pi \Phi}$.
The three-legged vertex $\Gamma^{\Phi \Phi \Pi} $ is represented by a black triangle with a 
red (gray) bar reflecting the symmetry with respect to the permutation of the external fields.
}
\label{fig_feynman_sing}
\end{figure}

Let us now attempt to calculate the effect of the cubic part ${\cal{H}}_3$ of the
spin-wave interactions on the magnon propagators.
Therefore it is convenient to formulate 
the perturbation theory in terms of an imaginary-time functional integral.
Formally, we simply have to replace 
the field operators $\Phi_{\bd{k}} $ and $\Pi_{\bd{k}}$ by
Hermitian fields $ \Phi_{\bd{k}} ( \tau )$ and $\Pi_{\bd{k}} ( \tau )$
depending on imaginary time $\tau$. We denote the corresponding
Fourier components in frequency space by
$\Phi_K$ and $\Pi_K$ where $K = ( \bd{k} , i \omega )$
is the collective label for the momentum $\bd{k}$ and the 
bosonic Matsubara frequency $ i \omega $.
The quadratic part ${\cal{H}}_2$ of our spin-wave Hamiltonian
given in Eq.~(\ref{eq:H2bb}) corresponds then to the following Gaussian Euclidean action,
 \begin{eqnarray}
 S_2 [ {\Phi}, {\Pi} ] & = & \frac{\beta}{2} \sum_K \Bigl[
 \Delta_{\bd{k}} {\Pi}_{-K } {\Pi}_K 
 + \frac{ \epsilon_{\bd{k}}^2}{\Delta_{\bd{k}}}
  {\Phi}_{ - K } {\Phi}_K  
 \nonumber
 \\
 & & 
+ ( \omega + i A_{\bd{k}}^- ) ( {\Phi}_{ - K }  {\Pi}_K - 
  {\Pi}_{ - K } {\Phi}_{ K } ) \Bigr].
 \hspace{7mm}
 \label{eq:S2def}
 \end{eqnarray}
From the Gaussian action (\ref{eq:S2def}) we obtain the free propagators,
 \begin{subequations}
 \begin{eqnarray}
\beta \langle \Phi_K \Phi_{ K^{\prime}} \rangle & = & \delta_{ \bd{k} , - \bd{k}^{\prime}}   \delta_{ \omega , - \omega^{\prime} }    G_0^{\Phi \Phi} ( K ),
 \label{eq:Gphiphi}
 \\
 \beta \langle \Pi_K \Pi_{ K^{\prime}} \rangle & = & 
\delta_{ \bd{k} , - \bd{k}^{\prime}}   \delta_{ \omega , - \omega^{\prime} } 
 G_0^{\Pi \Pi} ( K ),
 \label{eq:Gpipi}
 \\
\beta \langle \Phi_K \Pi_{ K^{\prime}} \rangle & = &  \delta_{ \bd{k} , - \bd{k}^{\prime}}   \delta_{ \omega , - \omega^{\prime} }   G_0^{\Phi \Pi} ( K ),
 \label{eq:Gphipi}
 \end{eqnarray}
 \end{subequations}
with
 \begin{subequations}
 \begin{eqnarray}
G_0^{\Phi \Phi} ( K ) & = & \frac{ 
  \Delta_{\bd{k}}}{ 
 \epsilon_{\bd{k}}^2   +  
 \left[ \omega +  i A_{\bd{k}}^- \right]^2 }
 \approx  \frac{ 
  \Delta_{0}}{ \epsilon_{\bd{k}}^2    +   \omega^2 } ,
 \label{eq:G0phiphi}
 \\
 G_0^{\Pi \Pi} ( K ) & = & \frac{ 
 \epsilon_{\bd{k}}^2 / \Delta_{\bd{k}} }{ 
 \epsilon_{\bd{k}}^2 +
 \left[ \omega +  i A_{\bd{k}}^-  \right]^2 }
 \approx  \frac{ 
  \epsilon_{\bd{k}}^2 / \Delta_{0}}{ \epsilon_{\bd{k}}^2    +   \omega^2 },
 \label{eq:G0pipi}
 \\
G_0^{\Phi \Pi} ( K ) & = & \frac{ 
  \omega +  i A_{\bd{k}}^-  
 }{  \epsilon_{\bd{k}}^2  +  
 \left[ \omega +  i A_{\bd{k}}^-  \right]^2 }
 \approx  \frac{ 
  \omega}{ \epsilon_{\bd{k}}^2    +   \omega^2 }.
 \label{eq:G0phipi}
 \end{eqnarray} 
 \end{subequations}
We have also given the leading approximations for small $\bd{k}$
where $\epsilon_{\bd{k}} \approx v ( \hat{\bd{k}} ) | \bd{k} |$.
In the presence of interactions the coefficients in Eq.~(\ref{eq:S2def})
acquire self-energy corrections, so that
the quadratic part of the true effective action is of the form
 \begin{eqnarray}
 \Gamma_2 [ {\Phi}, {\Pi} ] & = & \frac{\beta}{2} \sum_K \Bigl\{
 \left[ \Delta_{\bd{k}} + \Sigma^{\Pi \Pi} ( K ) \right] {\Pi}_{-K } {\Pi}_K
   \nonumber
 \\
 & + &
   \left[  \epsilon_{\bd{k}}^2 / \Delta_{\bd{k}}
 + \Sigma^{\Phi \Phi} ( K ) \right]  {\Phi}_{ - K } {\Phi}_K
 \nonumber
 \\
 & + &   \left[
    \omega +  i A_{\bd{k}}^-   + \Sigma^{\Phi \Pi} ( K ) \right] 
( {\Phi}_{ - K }   {\Pi}_K - 
 {\Pi}_{ - K }  {\Phi}_{ K } ) \Bigr\}  .
 \nonumber
 \\
 & &
 \end{eqnarray}
The corresponding propagators are
 \begin{subequations}
 \begin{eqnarray}
G^{\Phi \Phi} ( K ) & = & \frac{ 
  \Delta_{\bd{k}} + \Sigma^{\Pi \Pi} ( K )
 }{ D ( K ) },
 \\
 G^{\Pi \Pi} ( K ) & = & \frac{ 
 \epsilon_{\bd{k}}^2 / \Delta_{\bd{k}}
 + \Sigma^{\Phi \Phi} ( K ) }{ D( K )},
 \\
G^{\Phi \Pi} ( K ) & = & \frac{ 
  \omega +  i A_{\bd{k}}^-   + \Sigma^{\Phi \Pi} ( K )
 }{ D ( K ) },
 \end{eqnarray} 
 \end{subequations}
where the determinant is given by
 \begin{eqnarray}
 D( K ) & = & 
 \left[ \Delta_{\bd{k}} + \Sigma^{\Pi \Pi} ( K ) \right]
\left[ \epsilon_{\bd{k}}^2 / \Delta_{\bd{k}}
 + \Sigma^{\Phi \Phi} ( K ) \right] 
 \nonumber
 \\
 & + &  
\left[
    \omega +  i A_{\bd{k}}^-   + \Sigma^{\Phi \Pi} ( K ) \right]^2 .
 \end{eqnarray}
The crucial advantage of the above Hermitian field
parametrization is that it allows us to
isolate in a very simple way the infrared-divergent contributions to the self-energies
arising in second order perturbation theory in the cubic vertices.
 It turns out that only terms involving
the vertex
$\Gamma^{ \Phi \Phi \Pi }
 ( \bd{k}_1 , \bd{k}_2 ; \bd{k}_3 )  =    V_+ ( \bd{k}_3 )$
 give rise to
infrared-divergent terms. Moreover, since we are only interested in the infrared divergencies,
we may approximate
 \begin{equation}
  \Gamma^{ \Phi \Phi \Pi }
 ( \bd{k}_1 , \bd{k}_2 ; \bd{k}_3 )  \approx      V_+ ( 0 ) = 
\sin \theta \cos \theta \frac{h_c}{\sqrt{S}}.
 \label{eq:trimag}
 \end{equation}
The Feynman diagrams giving the leading singular contributions to the
self-energies are shown in Fig.~\ref{fig_feynman_sing}.
For small wave vectors these diagrams correspond to the
following self-energies,
 \begin{subequations}
 \begin{eqnarray}
 \Sigma^{\Phi \Phi} ( K ) & \approx & - \frac{  | V_+ ( 0 ) |^2}{ \beta N} 
\sum_{ K^{\prime}} \bigl[
 G_0^{\Phi \Phi } ( K^{\prime} ) G_0^{\Pi \Pi} ( K - K^{\prime} )
 \nonumber
 \\
 & & \hspace{12mm}
+
 G_0^{\Phi \Pi } ( K^{\prime} ) G_0^{\Pi \Phi} ( K - K^{\prime} )
\bigr] ,
 \label{eq:sigma1}
 \\
 \Sigma^{\Pi \Pi} ( K ) & \approx & - \frac{  | V_+ ( 0 ) |^2}{2 \beta N} 
\sum_{ K^{\prime}} G_0^{\Phi \Phi } ( K^{\prime} ) G_0^{\Phi \Phi} ( K - K^{\prime} ),
 \nonumber
 \\
 & &
 \label{eq:sigma2}
 \\
  \Sigma^{\Phi \Pi} ( K ) & \approx & - \frac{  | V_+ ( 0 ) |^2}{ \beta N} 
\sum_{ K^{\prime}} 
 G_0^{\Phi \Phi } ( K^{\prime} ) G_0^{\Pi \Phi} ( K - K^{\prime} ).
 \nonumber
 \\
 & &
 \label{eq:sigma3}
 \end{eqnarray}
\end{subequations}
Using the long-wavelength approximations 
of the Gaussian propagators given in Eqs.~(\ref{eq:G0phiphi})-(\ref{eq:G0phipi}), we obtain for small frequencies and
momenta to leading order,
  \begin{subequations}
 \begin{eqnarray}
 \Sigma^{\Phi \Phi} ( K ) & \approx & \alpha  \omega^2 / \Delta_0 ,
 \\
  \Sigma^{\Pi \Pi} ( K ) & \approx & -2 \alpha \Delta_0 ,
 \\
 \Sigma^{\Phi \Pi} ( K ) & \approx & - 2 \alpha \omega ,
 \end{eqnarray}
 \end{subequations}
where the dimensionless factor $ \alpha$ is given by
 \begin{equation}
 \alpha = \frac{ \Delta_0 | V_{+} ( 0 ) |^2}{16 N} \sum_{\bd{k}} \frac{1}{ \epsilon_{\bd{k}}^3} =
 \frac{ \cos \theta \sin^3 \theta }{16 S N} \sum_{\bd{k}} \frac{ h_c^3}{\epsilon_{\bd{k}}^3}.
 \end{equation}
Keeping in mind that for small wavevectors $\epsilon_{\bd{k}} \propto | \bd{k} |$,
it is obvious that
for $ D \leq 3$ the factor $\alpha$ 
is infrared divergent.
Note further that the first-order contribution generated by the quartic 
part of the Hamiltonian as given in Eq. (\ref{eq:H4hp}) is on
the same order in $1/S$ as the terms just discussed; however, the former contributions
are all finite and give rise to
finite renormalizations of the bare parameters, such as the spin-wave velocities $v_x$ and $v_y$.
We conclude that
the leading $1/S$ correction to linear spin-wave theory in the cone state of
Cs$_2$CuCl$_4$ diverges even at zero temperature.

\section{Correlation functions of 
interacting bosons in two dimensions}
\label{sec:bose}

In this section we show that
the infrared singularities in the $1/S$ expansion around the cone state of
Cs$_2$CuCl$_4$ are actually familiar from the theory of the condensed phase of the interacting Bose gas.
The linear spin-wave theory corresponds in the Bose gas to 
Bogoliubov's mean-field
theory for the condensed phase, leading to the well-known linear
phonon spectrum. But if one tries to  calculate fluctuation corrections
to mean-field theory one encounters in dimensions $D \leq 3$
infrared divergencies which arise from the
coupling between longitudinal and transverse fluctuations in the 
condensed phase.\cite{Shi98,Andersen04,Sinner09,*Sinner10,Dupuis11}
To leading order in perturbation theory, this coupling is described
by a triangular vertex similar to $ \Gamma^{\Phi\Phi\Pi}(\bd k_1, \bd k_2 ; \bd k_3)$
given in Eq.~(\ref{eq:vertexphiphipi}) where the field $\Pi$ 
describes  longitudinal fluctuations of the complex order parameter,
whereas the field $\Phi$ describes transverse fluctuations perpendicular
to the direction of the order parameter.
The reason why the  $1/S$-expansion of many other quantum magnets
is not plagued by similar singularities
is that usually the triangular vertex vanishes at long wavelengths due to spin 
conservation
or is identically zero by symmetry.
Conversely, if the magnon spectrum is gapless
and the usual $1/S$ expansion around the classical ground state 
generates a triangular vertex
which reduces to a constant at long wavelengths, then the leading 
$1/S$ correction to linear spin-wave theory is singular in dimensions $D \leq 3$.
Similar singularities appear also in the interaction
corrections to the spin-wave gap 
in anisotropic square lattice antiferromagnets.\cite{Maleyev00,Syromyatnikov01}
In this case the magnon spectrum is not gapless, but the triangular vertices 
in the $1/S$ expansion are finite, leading to a singular renormalization of the
spin-wave gap.

To obtain physically meaningful results,
the singularities encountered in the $1/S$ expansion around the cone state 
should be cured by some nonperturbative procedure which
takes the singular corrections to all orders in the $1/S$ expansion  into account.
Although within the conventional formulation of the $1/S$ expansion this cannot be performed,
we can achieve this by using known nonperturbative results for the interacting Bose 
gas\cite{Weichman88,Giorgini92,Castellani97,Pistolesi04}.
Let us therefore summarize in this section the relevant results for the Bose 
gas. In  Sec.~\ref{sec:spinstruc} we will then
apply these results to our model for Cs$_2$CuCl$_4$.

We consider a system of  interacting bosons confined to a volume $V$ with Hamiltonian,
 \begin{equation}
 \tilde{\cal{H}} = \sum_{\bd{k}} \frac{ \bd{k}^2}{2m} b^{\dagger}_{\bd{k}} 
 b_{\bd{k}} + \frac{1}{2V}
 \sum_{\bd{k} \bd{k}^{\prime} \bd{q}} u_{\bd{q}}
 b^{\dagger}_{\bd{k} + \bd{q}} b^{\dagger}_{\bd{k}^{\prime} - \bd{q}}
 b_{\bd{k}^{\prime} } b_{\bd{k}},
 \label{eq:Hb}
 \end{equation} 
where the Fourier transform $u_{\bd{q}}$ of the interaction has a finite limit for
vanishing momentum transfer $\bd{q}$.
In the condensed phase the single particle state with $\bd{k}=0$
is macroscopically occupied, so that the expectation value
$\langle b_{\bd{k} =0 } \rangle$ is proportional to $\sqrt{V}$.
We take this macroscopic occupation 
of the $\bd{k} =0$ state 
into account
via the usual Bogoliubov shift. 
In order to emphasize the analogy
with the spin-wave approach for   Cs$_2$CuCl$_4$
outlined in Sec.~\ref{sec:spinwave}, 
it is convenient to assume that the expectation value of $b_{\bd{k} =0}$ is 
purely imaginary so that we should set
 \begin{equation}
 b_0  =  i \sqrt{N_0} + \delta b_0,
 \label{eq:Bshift}
 \end{equation}
where $N_0$ is the number of condensed bosons.
Substituting this into Eq.~(\ref{eq:Hb}) and subtracting
the chemical potential term $\mu {\cal{N}} = \mu \sum_{\bd{k}}
 b^{\dagger}_{\bd{k}} b_{\bd{k}}$, 
we obtain
 \begin{equation}
  \tilde{\cal{H}} - \mu {\cal{N}} =   \left( \frac{\rho_0 u_0  }{2} - \mu \right)  N_0 +
 \tilde{\cal{H}}_1 + \tilde{\cal{H}}_2 + \tilde{\cal{H}}_3
  + \tilde{\cal{H}}_4,
 \end{equation}
where
 \begin{eqnarray}
 {\cal{\tilde{H}}}_1 & = &  i \sqrt{N_0} (  \rho_0 u_0  - \mu )  ( \delta b_0^{\dagger} - 
 \delta b_0),
 \\
 {\cal{\tilde{H}}}_2 & = & \sum_{\bd{k}} \Bigl\{ \Bigl[ \frac{ {\bd{k}}^2}{2m} 
 +  \rho_0 u_{\bd{k}} + \rho_0 u_0  - \mu \Bigr] b^{\dagger}_{\bd{k}} b_{\bd{k}}
 \nonumber
 \\
 & &  \hspace{7mm} - \frac{\rho_0 u_{\bd{k}} }{2}
 \bigl[ 
 b^{\dagger}_{\bd{k}} b^{\dagger}_{-\bd{k}} +
 b_{-\bd{k}} b_{\bd{k}} \bigr] \Bigr\},
 \\
{\cal{\tilde{H}}}_3  & = &  \frac{1}{\sqrt{N_0} }
 \sum_{ \bd{k}_1 \bd{k}_2 \bd{k}_3} \delta_{ \bd{k}_1 +
 \bd{k}_2 + \bd{k}_3  , 0 } \Bigl[
 \nonumber
 \\
 &  & \frac{i \rho_0}{ 2!}  ( u_{\bd{k}_1} + u_{\bd{k}_2} )
 b^{\dagger}_{  - \bd{k}_1 } b^{\dagger}_{  - \bd{k}_2 }
 b_{  \bd{k}_3 }  
 \nonumber
 \\
 & - & \frac{ i \rho_0 }{ 2!}  
  ( u_{\bd{k}_2} + u_{\bd{k}_3} )
 b^{\dagger}_{ - \bd{k}_1 } b_{  \bd{k}_2 }
 b_{ \bd{k}_3 }
\Bigr].
 \label{eq:H3bos}
 \end{eqnarray}
Here 
 $\rho_0 = N_0 / V $ is the condensate density. 
In ${\cal{\tilde{H}}}_2$ and  ${\cal{\tilde{H}}}_3$
it is understood that we should substitute
$b_{\bd{k} =0} \rightarrow \delta b_0$. With this convention
 ${\cal{\tilde{H}}}_4$ has the same form as the interaction
in our original Hamiltonian (\ref{eq:Hb}).
As in the spin-wave approach,
we demand that the linear term ${\cal{\tilde{H}}}_1$ vanishes identically,
implying the Hugenholtz-Pines identity,\cite{Shi98,Andersen04}
 \begin{equation} 
 \rho_0 u_0  = \mu ,
 \end{equation}
which fixes the condensate density $\rho_0 = \mu / u_0$ as a function of
the chemical potential and the  interaction.
The quadratic term then simplifies to
 \begin{align}
 {\cal{\tilde{H}}}_2 & =  \sum_{\bd{k}} \Bigl\{ \Bigl[ \frac{ {\bd{k}}^2}{2m} +  \rho_0 u_{\bd{k}}   \Bigr] b^{\dagger}_{\bd{k}} b_{\bd{k}}
 - \frac{\rho_0 u_{\bd{k}} }{2}
 \bigl[ 
 b^{\dagger}_{\bd{k}} b^{\dagger}_{-\bd{k}} +
 b_{-\bd{k}} b_{\bd{k}} \bigr] \Bigr\}.
 \label{eq:H2bos2}
 \end{align}
After diagonalization via a Bogoliubov transformation we obtain
\begin{equation}
 {\cal{\tilde{H}}}_2 = \sum_{\bd{k}} \left[ 
 {\epsilon}_{\bd{k}} \beta^{\dagger}_{\bd{k}} \beta_{\bd{k}} + \frac{1}{2} 
 \left(  {\epsilon}_{\bd{k}} -    \frac{ {\bd{k}}^2}{2m} - \rho_0 u_{\bd{k}}  \right) \right]\,,
 \label{eq:H2diagbos}
 \end{equation}
where $\beta^{\dagger}_{\bd{k}}$ and $\beta_{\bd{k}}$ are again canonical boson operators 
and the boson dispersion is
 \begin{eqnarray}
 {\epsilon}_{\bd{k}} 
 & =  & \sqrt{ \Bigl(   \frac{ {\bd{k}}^2}{2m} +  \rho_0 u_{\bd{k}}     \Bigr)^2 -  (\rho_0 u_{\bd{k}})^2 }
 \nonumber
 \\ 
 & = &  \sqrt{ \frac{ \bd{k}^2 }{m} \rho_0 u _{\bd{k}} + \left( 
 \frac{ \bd{k}^2}{2m} \right)^2}. 
 \label{eq:bosdispersion}
 \end{eqnarray}
For small wave vectors we obtain the well-known linear phonon dispersion,
 \begin{equation}
  \epsilon_{\bd{k}} = c_0 | \bd{k} | + {\cal{O}} ( \bd{k}^3 ),
 \end{equation}
with phonon velocity 
 \begin{equation}
 c_0 = \sqrt{ \frac{\rho_0 u_0 }{ m} } = \sqrt{ \frac{\mu}{m}}.
 \end{equation}
To facilitate the identification of the infrared divergent terms in perturbation theory,
we now introduce again Hermitian field operators
 as in Eq.~(\ref{eq:hermit}).  Note that with our phase convention
the $\Pi$ field describes longitudinal fluctuations in the direction of the order
parameter
whereas the $\Phi$ field is associated with transverse fluctuations.
Substituting the transformation (\ref{eq:hermit}) into
Eq.~(\ref{eq:H2bos2}) we obtain for the quadratic part of the Hamiltonian,
 \begin{eqnarray}
 {\cal{\tilde{H}}}_2 & = & \frac{1}{2} \sum_{\bd{k}}
 \Bigl[ \Bigl( 2 \rho_0 u_{\bd{k}} + \frac{ \bd{k}^2}{2m} \Bigr)
  {\Pi}_{- \bd{k} } {\Pi}_{\bd{k}}  + \frac{ \bd{k}^2}{2m} 
  {\Phi}_{ - \bd{k} } {\Phi}_{\bd{k}} 
 \nonumber
 \\
  & &  \hspace{10mm} - 
 \Bigl(  \rho_0 u_{\bd{k}} + \frac{ \bd{k}^2}{2m} \Bigr)
 \Bigr]
 \nonumber
 \\
 & = &  \frac{1}{2} \sum_{\bd{k}} \Bigl[
 \Delta_{\bd{k}} {\Pi}_{- \bd{k} } {\Pi}_{\bd{k}} 
 + \frac{ \epsilon_{\bd{k}}^2}{ \Delta_{\bd{k}}}
  {\Phi}_{ - \bd{k} } {\Phi}_{\bd{k}}
-  \Bigl(  \rho_0 u_{\bd{k}} + \frac{ \bd{k}^2}{2m} \Bigr)  \Bigr],
 \nonumber
 \\
 & &
 \label{eq:hbos2}
 \end{eqnarray}
where in this section,
 \begin{equation}
  \Delta_{\bd{k}} =  2 \rho_0 u_{\bd{k}} + \frac{ \bd{k}^2}{2m} .
 \end{equation}
Our notation emphasizes  the formal similarity between Eqs.~(\ref{eq:hbos2})
and (\ref{eq:H2swphi}); the additional term in Eq.~(\ref{eq:H2swphi}) 
involving the antisymmetric factor $A_{\bd{k}}^{-}$ complicates the expression,
but does not change the final result because of Eq.~(\ref{eq:Aminussmall}).
Ignoring again commutator terms involving 
a single power of the fields,
the leading interaction part  ${\cal{\tilde{H}}}_3$ of our boson Hamiltonian
can be written as
 \begin{eqnarray}
 \tilde{{\cal{H}}}_3 & = & \frac{1}{\sqrt{N_0}} 
 \sum_{ \bd{k}_1 \bd{k}_2 \bd{k}_3} \delta_{ \bd{k}_1 +
 \bd{k}_2 + \bd{k}_3  , 0 } \Bigl[
 \nonumber
 \\
 &  & 
\frac{1}{ 3!}
\Gamma^{ \Pi \Pi \Pi }
 ( \bd{k}_1 , \bd{k}_2 , \bd{k}_3 )
  {\Pi}_{   \bd{k}_1 } {\Pi}_{   \bd{k}_2 }
 {\Pi}_{  \bd{k}_3 }  
 \nonumber
 \\
 & + & 
\frac{1}{ 2!}
\Gamma^{ \Phi \Phi \Pi }
 ( \bd{k}_1 , \bd{k}_2 ; \bd{k}_3 )
   {\Phi}_{   \bd{k}_1 } {\Phi}_{   \bd{k}_2 }
  {\Pi}_{  \bd{k}_3 } \Bigr] ,
 \label{eq:H3boshermitian}
 \end{eqnarray}
where the symmetrized vertices are
\begin{subequations}
 \begin{align}
  \Gamma^{\Pi\Pi\Pi}(\bd k_1,\bd k_2,\bd k_3)=&  \sqrt{2} \rho_0 
 \left(
 u_{\bd k_1} + u_{\bd{k}_2} +u_{\bd k_3} \right),\\
  \Gamma^{\Phi\Phi\Pi}(\bd k_1, \bd k_2 ; \bd k_3)=& \sqrt{2} \rho_0 u_{\bd k_3}\,.
 \label{eq:vertex3bos}
 \end{align}
\end{subequations}

As in Sec.~\ref{subsec:hermit} we use the functional integral formulation
of the problem.  We are interested in the bosonic two-point functions
$G^{\Phi \Phi} ( K )$, $G^{\Pi \Pi} ( K)$, and $G^{\Phi \Pi} ( K )$,
which are defined in terms of functional averages as in Eqs.~(\ref{eq:Gphiphi})-(\ref{eq:Gphipi}).
At long wavelengths, the Gaussian approximation is formally identical with 
Eqs.~(\ref{eq:G0phiphi})-(\ref{eq:G0phipi}) where now
 $\Delta_0 = 2 \rho_0 u_0$ and $\epsilon_{\bd{k}} = c_0 | \bd{k} |$.
However,  as pointed out in Sec.~\ref{subsec:hermit},
the linear phonon dispersion in combination with the finite 
limit of the cubic vertex,
 \begin{equation}
 \Gamma^{\Phi\Phi\Pi}(0, 0 ; 0) = \sqrt{2} \rho_0 u_0
 \label{eq:tribos}
 \end{equation}
give rise to infrared divergencies in 
perturbation theory for all dimensions $D \leq 3$.
Fortunately, nonperturbative resummations 
 of these divergencies are available for the interacting 
Bose gas\cite{Weichman88,Castellani97,*Pistolesi04,Sachdev99b,Zwerger04,Kreisel07,Dupuis11}
so that we know the true infrared behavior of the above two-point functions.
Using the expressions derived by Castellani {\it{et al.}}\cite{Castellani97} we obtain
for small momenta and frequencies in two dimensions,\cite{Kreisel07}
 \begin{subequations}
  \begin{align}
  G^{\Phi \Phi} ( K ) & =  \frac{ \Delta}{ c^2 \bd{k}^2 + \omega^2 },
 \label{eq:Gphiphitrue}
 \\
   G^{\Pi \Pi} ( K ) & = 
 \frac{ \Delta^2 }{  8 \pi^2 \rho_0 c^2  } \frac{ 1}{
 \sqrt{ c^2    {\bd{k}}^2    +   \omega^2} } 
 -  \frac{   Z_{\parallel}^2  \omega^2  / \Delta }{  c^2 {\bd{k}}^2  + \omega^2}
  \; ,
 \label{eq:Gpipitrue}
 \\
  G^{\Phi \Pi} ( K ) & =   \frac{   Z_{\parallel}   \omega }{ c^2 \bd{k}^2 + \omega^2 },
 \label{eq:Gphipitrue}
 \end{align}
 \end{subequations}
where the dimensionless factor $Z_{\parallel}$ can be expressed in terms of the
derivative of the condensate density $\rho_0$ with respect to the chemical potential
as,\cite{Castellani97,Kreisel07}
 \begin{equation}
 Z_{\parallel} = \frac{ \Delta}{2 \rho_0} \frac{ d \rho_0}{d \mu }.
 \end{equation}
The important point is that the longitudinal correlation function
(\ref{eq:Gpipitrue}) contains a nonanalytic contribution 
which has first been discussed by Weichman.\cite{Weichman88}
The above infrared behavior of the correlation functions
involves three independent parameters: the renormalized sound velocity $c$,
the energy scale $\Delta$, and the true condensate density $\rho_0$.
These parameters can in principle be calculated perturbatively or numerically as functions
of the bare parameters $m$, $\mu$, and $u_0$
of the model defined in Eq.~(\ref{eq:Hb}).
At the level of the Gaussian approximation we obtain
 \begin{subequations}
 \begin{eqnarray}
 c & \approx & c_0 = \sqrt{ \rho_0 u_0/m } 
= \sqrt{ {\mu}/{m}},
 \\
 \Delta & \approx & \Delta_0 = 2 \rho_0 u_0 = 2 m c_0^2 = 2 \mu,
 \\
\rho_0 & \approx  &  \mu /u_0,
 \end{eqnarray}
\end{subequations}
implying 
 $Z_{\parallel}  \approx 1$.
In general, $Z_{\parallel}$ as well as 
the ratios $ c / c_0 \equiv Z_c $ and $\Delta / \Delta_0$ deviate from unity.
Following Ref.~[\onlinecite{Castellani97}], it is convenient to introduce the
renormalization factor $Z_{\rho} = \rho / \rho_0$,
where $\rho$ is the total density of the bosons. 
Then  a Ward identity implies that $\Delta / \Delta_0 = Z_c^2 / Z_{\rho}$.
The three dimensionless renormalization factors  $Z_c$, $Z_{\rho}$, and
$Z_{\parallel}$ must be fixed from microscopic calculations or from
experiments.
Finally, let us emphasize that the nonanalytic part of $G^{\Pi \Pi} ( K )$
in Eq.~(\ref{eq:Gpipitrue}) becomes important for
wavevectors $| \bd{k} |$ smaller than the Ginzburg scale $k_G$,
which is in two dimensions given by\cite{Kreisel07}
 \begin{equation}
 k_G \approx \frac{ (mc)^3}{\rho_0} \approx \frac{ \Delta^3}{ 8 \rho_0 c^3}.
 \end{equation}

\section{Spin structure factor in the cone state}
\label{sec:spinstruc}

Due to the structural similarity between the
above results for the interacting Bose gas and the
infrared behavior of spin-wave theory in the
cone state of Cs$_2$CuCl$_4$ developed in Sec.~\ref{subsec:hermit},
we may use the non-perturbative  results (\ref{eq:Gphiphitrue})-(\ref{eq:Gphipitrue})
to calculate the two-point functions of the spin-wave excitations in
Cs$_2$CuCl$_4$.  A slight mismatch between the two theories is due to the fact
that
the spin-wave spectrum in Cs$_2$CuCl$_4$ is anisotropic,
whereas in our Bose gas Hamiltonian (\ref{eq:Hb})
we have assumed an isotropic dispersion.
To obtain a mapping between these two models,
we simply use the angular average of the
direction-dependent magnon velocity $v ( \hat{\bd{k}} )$
given in Eqs.~(\ref{eq:vkhat})-(\ref{eq:vy}),
 \begin{equation}
 \bar{v} = \int_{0}^{2 \pi} \frac{ d \varphi}{2 \pi} \sqrt{ v_x^2 \cos^2 \varphi + v_y^2
 \sin^2 \varphi }.
 \label{eq:angular}
 \end{equation} 
In Fig.~\ref{fig:velocities}, we show the average velocity $\bar{v}$ together with
$v_x$ and $v_y$ as a function of the magnetic field.
To leading order in the $1/S$ expansion we should then identify
in Eqs.~(\ref{eq:Gphiphitrue})-(\ref{eq:Gphipitrue}),
 \begin{subequations}
 \begin{eqnarray}
 c & \approx & \bar{v}  = \bar{v}_{0}  \sin \theta = \bar{v}_0 \sqrt{ 1 - (h/h_c)^2}
 \nonumber
 \\
 & \approx &  \bar{v}_0 \sqrt{2}   \sqrt{ 1 - h/h_c} ,
 \label{eq:csub}
 \\
 \Delta & \approx & h_c \sin^2 \theta  = h_c [ 1 - (h/h_c)^2 ]
 \nonumber
 \\
& \approx & 2 ( h_c - h ),
 \label{eq:deltasub}
 \\
 \rho_0 & \approx & \frac{nS}{2} \tan^2 \theta = \frac{nS}{2} [ (h_c/h)^2 -1 ]
 \nonumber
 \\
 & \approx & 
  n S (1 - h/ h_c ).
 \label{eq:rho0id}
 \end{eqnarray}
 \end{subequations}
Here $\bar{v}_0$ is the average spin-wave velocity for $h=0$ 
where the spins form a spiral in the $x$-$y$-plane and
$n = N/ V$ is the number of spins per unit volume. 
The approximate equalities in Eqs.~(\ref{eq:csub})-(\ref{eq:rho0id})
are valid if $h$ is slightly smaller than the saturation field $h_c$;
only in this regime is our mapping between the spin system
and the Bose gas accurate.
The identification (\ref{eq:rho0id}) follows from the requirement that
the triangular vertex $\Gamma^{\Phi \Phi \Pi} (0,0;0 )/ \sqrt{N_0}$
in the Bose gas given in Eq.~(\ref{eq:tribos})
should be equal to the corresponding vertex
 $\Gamma^{\Phi \Phi \Pi} (0,0;0 )/ \sqrt{N}$
in the magnon gas given in Eq.~(\ref{eq:trimag}).
Obviously, for  $0 < h_c - h  \ll h_c$ 
the difference $h_c - h $
is analogous to the chemical potential in the Bose gas,
whereas $h_c / (nS)$ corresponds to 
the interaction $u_0$ at vanishing momentum transfer.
Since the interaction vertices in the spin system 
are proportional to increasing powers of $1/S$, for large $S$
all renormalization factors $Z_c$, $Z_{\rho}$, and $Z_{\parallel}$ 
approach unity.
Although for $S=1/2$ these factors are expected to deviate significantly from
unity, we will not attempt to calculate these corrections here.
But we can implicitly take these renormalization factors into account
by fixing the unknown parameters $\bar{v}$, $\Delta$, and $\rho_0$ from experiments.
Therefore, we use the relations (\ref{eq:csub})-(\ref{eq:rho0id}) but
substitute experimental values for the average spin-wave velocity $\bar{v}$,
the critical field $h_c$, and saturated spin density $s$.
Because the nonanalytic contribution to the longitudinal correlation function (\ref{eq:Gpipitrue})
cannot be obtained in any finite order perturbation theory,
our approach based on the mapping
to the Bose gas 
effectively resums the singular terms in the spin-wave expansion
to all orders in $1/S$.

To make contact with neutron-scattering experiments,
we need the  spin dynamic structure factor 
in the cone state of Cs$_2$CuCl$_4$,
which is defined by
 \begin{equation}
 S^{\alpha \beta} ( \bd{k} , \omega ) = \int_{ - \infty}^{\infty} \frac{dt}{2 \pi}  e^{ i \omega t } 
 \langle S^{\alpha}_{- \bd{k}} ( t ) S^{\beta}_{\bd{k}} ( 0 ) 
 \rangle,
 \label{eq:Sdyndef}
 \end{equation}
where $\alpha , \beta = a,b,c$ label the various crystallographic axes
and the Fourier components of the spin operators are defined by
 \begin{equation}
   \bd{S}_{\bd{k}} = \frac{1}{\sqrt{N}} \sum_i e^{ -i \bd{k} \cdot {\bd{R}}_i } {\bd{S}}_i.
 \end{equation}
Previously, Veillette {\it{et al.}}\cite{Veillette05}
have calculated  $S^{\alpha \beta} ( \bd{k} , \omega ) $
in the ground state of Cs$_2$CuCl$_4$ 
for a vanishing magnetic field within spin-wave theory. They found that spin-wave interactions
give rise to extended scattering continua in the spectral lineshape.
Note that for $h=0$ the triangular vertex $\Gamma^{\Phi \Phi \Pi} (0,0;0)$ vanishes 
so that in the planar spiral state
there are no singular terms in the $1/S$ expansion.
We are not aware of any calculations 
analogous to those of Ref.~[\onlinecite{Veillette05}]
for a finite magnetic field.
From Sec.~\ref{subsec:hermit} it is clear that in this case
some $1/S$ corrections are infrared divergent.  
Using the nonperturbative results
(\ref{eq:Gphiphitrue})-(\ref{eq:Gphipitrue}) for the interacting Bose gas,
we can now resum these divergencies and determine the
associated spectral lineshape.

In order to calculate the spin structure factor (\ref{eq:Sdyndef}), 
we note that within linear spin-wave theory 
our Hermitian operators  $\Phi_{\bd{k}}$ and $\Pi_{\bd{k}}$
defined in Eq.~(\ref{eq:hermit}) are simply related to the
projections of the spin operators onto the  local coordinate system
formed by the orthogonal triad $\bd{e}_i^{(1)}$,  $\bd{e}_i^{(2)}$, 
and $\hat{\bd{m}}_i$ defined in Eqs.~(\ref{eq:ei1}, \ref{eq:ei2}, and \ref{eq:spiral}),
\begin{subequations}
 \begin{align}
 S_{\bd k}^{(1)} & = 
 \frac{1}{\sqrt{N}} \sum_i   \bd{{e}}_i^{(1) }
 \cdot \bd{S}_i   e^{-i\bd k \cdot {\bd{R}}_i}  \approx
  \sqrt{S} \Phi_{\bd k},
 \\
 S_{\bd k}^{(2)} & =  \frac{1}{\sqrt{N}} \sum_i   \bd{{e}}_i^{(2) }
 \cdot \bd{S}_i   e^{-i\bd k \cdot {\bd{R}}_i}  \approx
  \sqrt{S} \Pi_{\bd k},
 \\
 S_{\bd k}^\parallel &= 
\frac{1}{\sqrt{N}} \sum_i   \hat{\bd{m}}_i 
 \cdot \bd{S}_i   e^{-i\bd k \cdot {\bd{R}}_i} \notag\\
 &= S\sqrt{N}\delta_{\bd k,0} 
 -\frac{1}{2\sqrt{N}}\sum_{\bd q} \left[ \Phi_{-\bd q}\Phi_{\bd q+\bd k}+
 \Pi_{-\bd q}\Pi_{\bd q+\bd k} \right. 
 \notag\\
 & \left. \hspace{3mm}
 + i\Phi_{-\bd q}\Pi_{\bd q+\bd k} -i\Pi_{-\bd q}\Phi_{\bd q+\bd k}\right].
 \end{align}
\end{subequations}
Substituting these expressions into Eq.~(\ref{eq:Sdyndef}) and
using the analytic continuation of the nonperturbative results 
 (\ref{eq:Gphiphitrue})-(\ref{eq:Gphipitrue}) for real frequencies, we find that
in the local coordinate system the two transverse components $S^{(1)}_{\bd{k}}$ and
$S^{(2)}_{\bd{k}}$ give rise to the following contributions
to the dynamic structure factor for $\omega > 0$:
\begin{subequations}
 \begin{eqnarray}
  S^{11} (\bd k,\omega) & = & 
 - \frac{S}{\pi} \mathop{\mathrm{Im}} G^{\Phi \Phi} ( \bd{k} , \omega + i 0 ) 
 \nonumber
 \\
 & = &
\frac{S\Delta}{2 \omega} \delta(\omega-c|\bd k|),
 \label{eq:S11}\\
  S^{22}(\bd k,\omega) & = & 
 - \frac{S}{\pi} \mathop{\mathrm{Im}} G^{\Pi \Pi} ( \bd{k} , \omega + i 0 ) 
 \nonumber
 \\
 &  &  \hspace{-20mm} =
 \frac{ S \Delta^2}{ (2 \pi )^3 \rho_0 c^2}
 \frac{  \Theta(\omega-c|\bd k|) }{ \sqrt{ \omega^2 - c^2 \bd{k}^2 }} 
+  \frac{ S \omega}{2 \Delta}  Z_{\parallel}^2 
 \delta(\omega-c|\bd k|),
 \label{eq:S22}
 \hspace{7mm}
\\
  S^{12}(\bd k,\omega) & = & - S^{21} ( \bd{k} , \omega ) =
 i \frac{S}{\pi} \mathop{\mathrm{Re}}  G^{\Phi \Pi} ( \bd{k} , \omega + i 0 ) 
 \nonumber
 \\ 
 & = & 
 i\frac{SZ_\parallel}{2} \delta(\omega-c|\bd k|)
 \label{eq:S12}.
 \end{eqnarray}
\end{subequations}
To this order in $S$, the component
$S^{\parallel}_{\bd k}$ of the spin operator parallel to the local magnetization
does not contribute to the 
inelastic part of the dynamical structure factor.

\begin{figure}
\includegraphics[width=\linewidth]{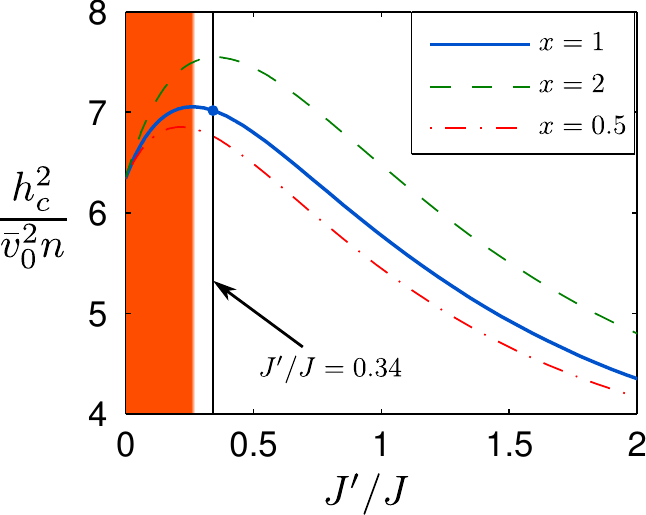}
\caption{(Color online) 
Plot of the dimensionless factor $h_c^2/n\bar v^2$  
appearing in Eqs.~(\ref{eq:factor1}), and(\ref{eq:factor2})
as a function of the ratio 
$J'/J$. To illustrate the dependence on the 
Dzyaloshinskii-Moriya interaction $D$ we have scaled 
$D$ with $J'$ such that at the experimental value of the ratio $J'/J$ 
(indicated by a vertical line) the experimental value $D_{\text{ex}}$ is given by
$D_{\text{ex}} = D / x $. The different curves
correspond to different values of $x$ as indicated in the caption, and
the dot marks the experimental values of $J$, $J'$ and $D$.  Quantum fluctuations
destroy the long-range order in the triangular lattice antiferromagnet in the shaded area
($J'/J\lesssim 0.27$), see Ref.~[\onlinecite{Trumper99}].
}
\label{fig:factor}
\end{figure}

\begin{figure}
\includegraphics[width=\linewidth]{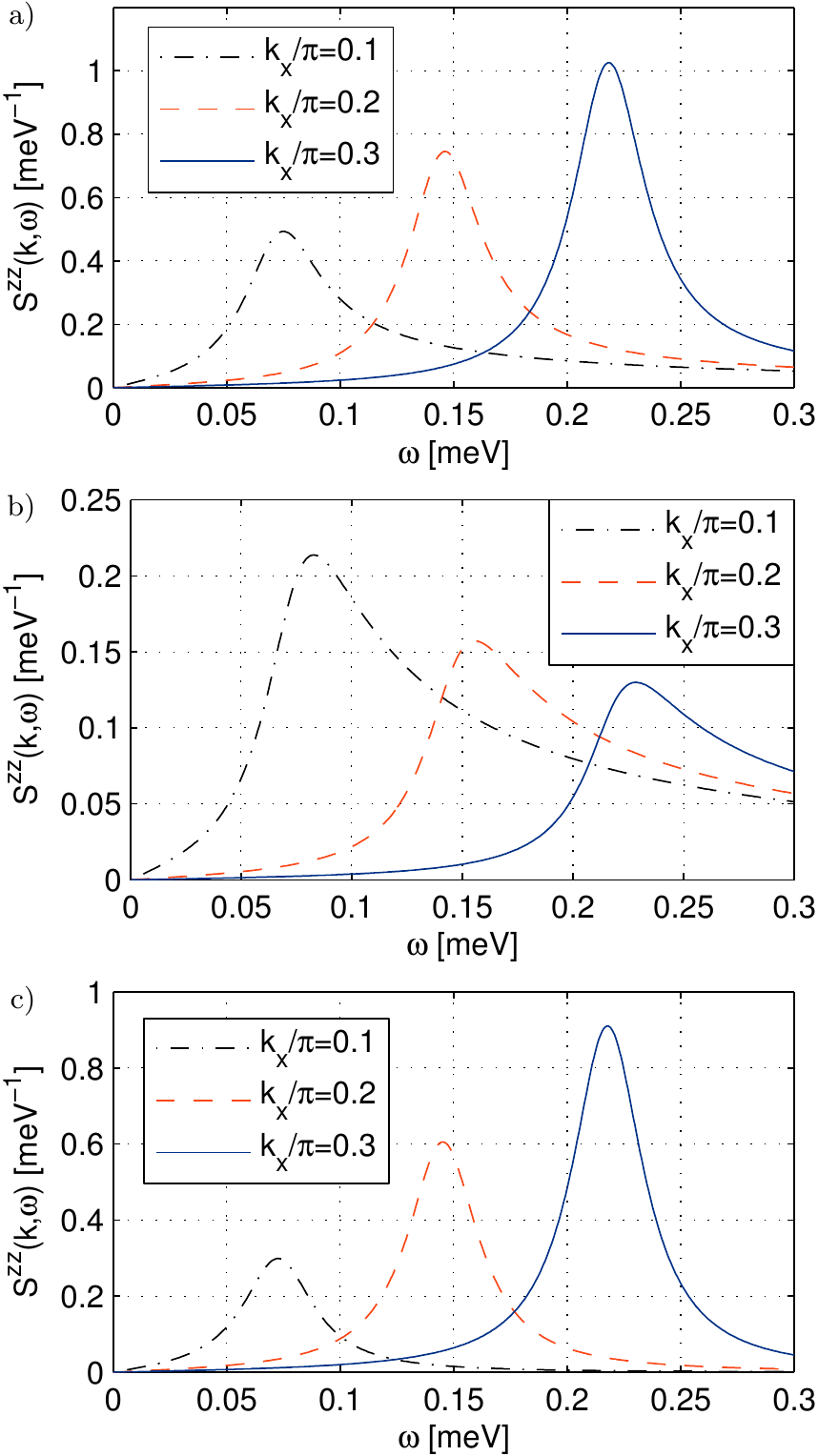}
\caption{(Color online) 
Graph of the component $S^{zz}(\bd k,\omega)$ of the dynamic structure factor
in the cone state of Cs$_2$CuCl$_4$ along the crystallographic $a$ axis for
a magnetic field of magnitude $h = 0.7 h_c$ parallel to the $a$ axis.
The curves are for $k_y=0$ and different values of $k_x$
as indicated in the captions.
The upper panel (a) shows the total contribution,
the middle panel (b) shows the anomalous contribution
given by the first term on the right-hand side of
Eq.~(\ref{eq:Szzt}), and the lower panel (c)
represents the $\delta$-function contribution given by the
last term in Eq.~(\ref{eq:Szzt}).
To take into account the
typical energy resolution in  neutron scattering experiments\cite{Coldea03}
we have convoluted the right-hand side of Eq.~(\ref{eq:Szzt})
with a 
Lorentzian of width $\gamma = 0.019\,\text{meV}$ (full width at half maximum). 
To fix the parameters $c$, $\Delta$, and $\rho_0$ appearing
in Eq.~(\ref{eq:Szzt}), we use the large-$S$ relations  (\ref{eq:csub})-(\ref{eq:rho0id}) with
$\bar{v}_0 =374\,\text{m/s}$ (compare Fig.~\ref{fig:velocities}) and $n=2/(bc)$ (see Fig.~\ref{fig_model}). For simplicity we have set $Z_{\parallel }=1$.
}
\label{fig:szz}
\end{figure}

To obtain the dynamic structure factor in the laboratory basis,
we express the Cartesian components of the spin operator
in terms of the components in the tilted basis.
Using the expansion (\ref{eq:Sexpansion}) and the definition of the tilted
basis given in Eqs.~(\ref{eq:spiral}, \ref{eq:ei1}, and \ref{eq:ei2})
we obtain for the Fourier components of the spin operators,
\begin{subequations}
 \begin{align}
  S_{\bd k}^x=&\frac{1}{2i}\left[ S_{\bd k-\bd Q}^{(1)}-S_{\bd k+\bd Q}^{(1)}\right]
 +\frac{\cos \theta}{2}\left[S_{\bd k-\bd Q}^{(2)}-S_{\bd k+\bd Q}^{(2)}\right]\notag\\
 &+ \frac{\sin \theta}{2}\left[ S_{\bd k-\bd Q}^{\parallel}+S_{\bd k+\bd Q}^\parallel\right],
 \\
  S_{\bd k}^y =&- \frac{1}{2}\left[ S_{\bd k-\bd Q}^{(1)}+S_{\bd k+\bd Q}^{(1)}\right]
 +\frac{\cos \theta}{2i}\left[ S_{\bd k-\bd Q}^{(2)}+S_{\bd k+\bd Q}^{(2)}\right] \notag\\
&+\frac{ \sin \theta}{2i}\left[ S_{\bd k-\bd Q}^{\parallel}-S_{\bd k+\bd Q}^\parallel\right],\\
  S_{\bd k}^z=&- \sin \theta S_{\bd k}^{(2)}+ \cos \theta S_{\bd k}^\parallel.
 \end{align}
\end{subequations}
This yields for the diagonal components of the dynamic structure factor in the laboratory basis,
 \begin{subequations}
 \begin{eqnarray}
  S^{xx}(\bd k,\omega)  & = & S^{yy}(\bd k,\omega)
 \nonumber
 \\
 &   & \hspace{-12mm} =
\frac{1}{4}
 \left[  S^{11}(\bd k+\bd Q,\omega) +S^{11}(\bd k-\bd Q,\omega) \right]
 \notag
 \\
 & &  \hspace{-12mm} + \frac{\cos^2 \theta}{4} \left[ S^{22}(\bd k+\bd Q,\omega)
 + S^{22}(\bd k-\bd Q,\omega)\right]
 \notag
 \\
 & &  \hspace{-12mm} + i \frac{ \cos \theta}{2} \left[
 S^{12}(\bd k-\bd Q,\omega)-S^{12}(\bd k+\bd Q,\omega)\right],
 \hspace{7mm}
 \label{eq:Sxx}
 \\
 S^{zz}(\bd k,\omega)  & = & \sin^2\theta\, S^{22}(\bd k,\omega).
 \label{eq:Szz}
 \end{eqnarray}
 \end{subequations}
For completeness, we also give the off-diagonal components,
 \begin{subequations}
 \begin{eqnarray}
  S^{xy}(\bd k,\omega)&= & - S^{yx}(\bd k,\omega)
 \notag
 \\
 &  & \hspace{-15mm} =
 \frac{ i}{ 4} \left[S^{11}(\bd k+\bd Q,\omega)+S^{11}(\bd k-\bd Q,\omega)\right] 
 \notag\\
 &  & \hspace{-12mm} + i \frac{\cos^2 \theta}{4} 
 \left[ S^{22}(\bd k+\bd Q,\omega)+S^{22}(\bd k-\bd Q,\omega) \right] 
 \notag\\
 & & \hspace{-12mm}  + \frac{  \cos \theta  }{2}  \left[ 
 S^{12}(\bd k+\bd Q,\omega)+S^{12}(\bd k-\bd Q,\omega)\right],
 \hspace{7mm}
 \\
 S^{xz}(\bd k,\omega)&= & - S^{zx}(\bd k,\omega ) = 0,
 \\
 S^{yz}(\bd k,\omega) & = & - S^{zy}(\bd k,\omega)=0.
 \end{eqnarray}
 \end{subequations}
Recall that we have chosen the $z$ direction such that it
agrees with the crystallographic $a$ axis, whereas the $x$ and $y$ directions
are associated with the $b$ and $c$ axes.
Substituting our nonperturbative
expressions for the components of the dynamic structure factor
in the tilted basis given in Eqs.~(\ref{eq:S11})-(\ref{eq:S12})
into Eqs.~(\ref{eq:Sxx}) and (\ref{eq:Szz}), we finally obtain
 \begin{eqnarray}
  & &  S^{xx}(\bd k,\omega)   =  S^{yy}(\bd k,\omega)
 = \frac{ S \cos^2 \theta  \Delta^2}{4 (2 \pi )^3 \rho_0 c^2 }
 \nonumber
 \\
 &  & \hspace{5mm} \times 
 \left[ 
 \frac{  \Theta(\omega-c|\bd k + \bd Q |) }{ \sqrt{ \omega^2 - c^2 ( \bd{k} + \bd{Q} )^2 }}
 +
 \frac{  \Theta(\omega-c|\bd k - \bd Q |) }{ \sqrt{ \omega^2 - c^2 ( \bd{k} - \bd{Q} )^2 }}
 \right] 
 \nonumber
 \\
 &    & + \frac{S \Delta}{8 \omega } \left[ 1 + \cos \theta Z_{\parallel} \frac{\omega}{\Delta}
 \right]^2 \delta(\omega-c|\bd k+\bd Q|)
 \nonumber
 \\
 &  & + \frac{S \Delta}{8 \omega } \left[ 1 - \cos \theta Z_{\parallel} \frac{\omega}{\Delta}
 \right]^2 \delta(\omega-c|\bd k - \bd Q|),
 \label{eq:Sxxyy}
 \end{eqnarray}
and for the  $zz$ component,
 \begin{eqnarray}
  S^{zz}(\bd k,\omega) & = & 
 \frac{ S \sin^2 \theta \Delta^2}{ (2 \pi )^3 \rho_0 c^2}
 \frac{  \Theta(\omega-c|\bd k|) }{ \sqrt{ \omega^2 - c^2 \bd{k}^2 }} 
 \notag
 \\
 & + &
  \frac{ S \sin^2 \theta \;  \omega}{2 \Delta}  Z_{\parallel}^2 
 \delta(\omega-c|\bd k|).
 \label{eq:Szzt}
 \end{eqnarray}
Using Eqs.~(\ref{eq:csub})-(\ref{eq:rho0id}) to estimate the quantities
$c$, $\Delta$ and $\rho_0$, we find that the
dimensionless prefactor of the nonanalytic continuum contribution 
to the transverse part $S^{xx} ( \bd{k} , \omega ) = S^{yy} ( \bd{k} , \omega ) $ 
of the structure factor can be written as
 \begin{equation}
 \frac{ S \cos^2 \theta  \Delta^2}{4 (2 \pi )^3 \rho_0 c^2 }
 \approx \frac{ \cos^4 \theta }{2 (2 \pi)^3} \frac{ h_c^2}{ n \bar{v}_0^2 },
 \label{eq:factor1}
 \end{equation}
which is maximal close to the saturation field.
On the other hand,
the corresponding prefactor in  the longitudinal structure factor $S^{zz} ( \bd{k} , \omega ) $ is
 \begin{equation}
 \frac{ S \sin^2 \theta \Delta^2}{ (2 \pi )^3 \rho_0 c^2}
 \approx \frac{  \sin^2(2 \theta)  }{2 (2 \pi)^3} \frac{ h_c^2}{ n \bar{v}_0^2 },
 \label{eq:factor2}
 \end{equation}
which has a maximum for $\theta = \pi/4$, corresponding to $h = h_c/\sqrt{2}$.
In Fig.~\ref{fig:factor} we
plot the dimensionless factor $h_c^2/( n \bar{v}_0^2 )$
as a function for $J^{\prime} / J$.
Obviously, this factor is maximal if $J^{\prime}/J$ is close to
the experimental value $J^{\prime}/ J \approx 0.34$.
Of course, our mapping between the spin system and the Bose gas
is only quantitatively accurate if $h$ is close to $h_c$, 
although for smaller $h$ the 
qualitative behavior of the dynamic structure factor
should still be given by the above expressions.
A plot  $S^{zz} ( \bd{k} , \omega ) $ for $h =0.7 h_c$ and for values for $c$ and 
$\Delta$ relevant for Cs$_2$CuCl$_4$ is shown in
Fig.~\ref{fig:szz}.
The pronounced asymmetry of the spectral line-shape factor shown in the
upper panel is due to the threshold singularity of the anomalous contribution
shown in the middle panel (b) of Fig.~\ref{fig:szz}. 
\begin{figure}[tb]
\includegraphics[width=\linewidth]{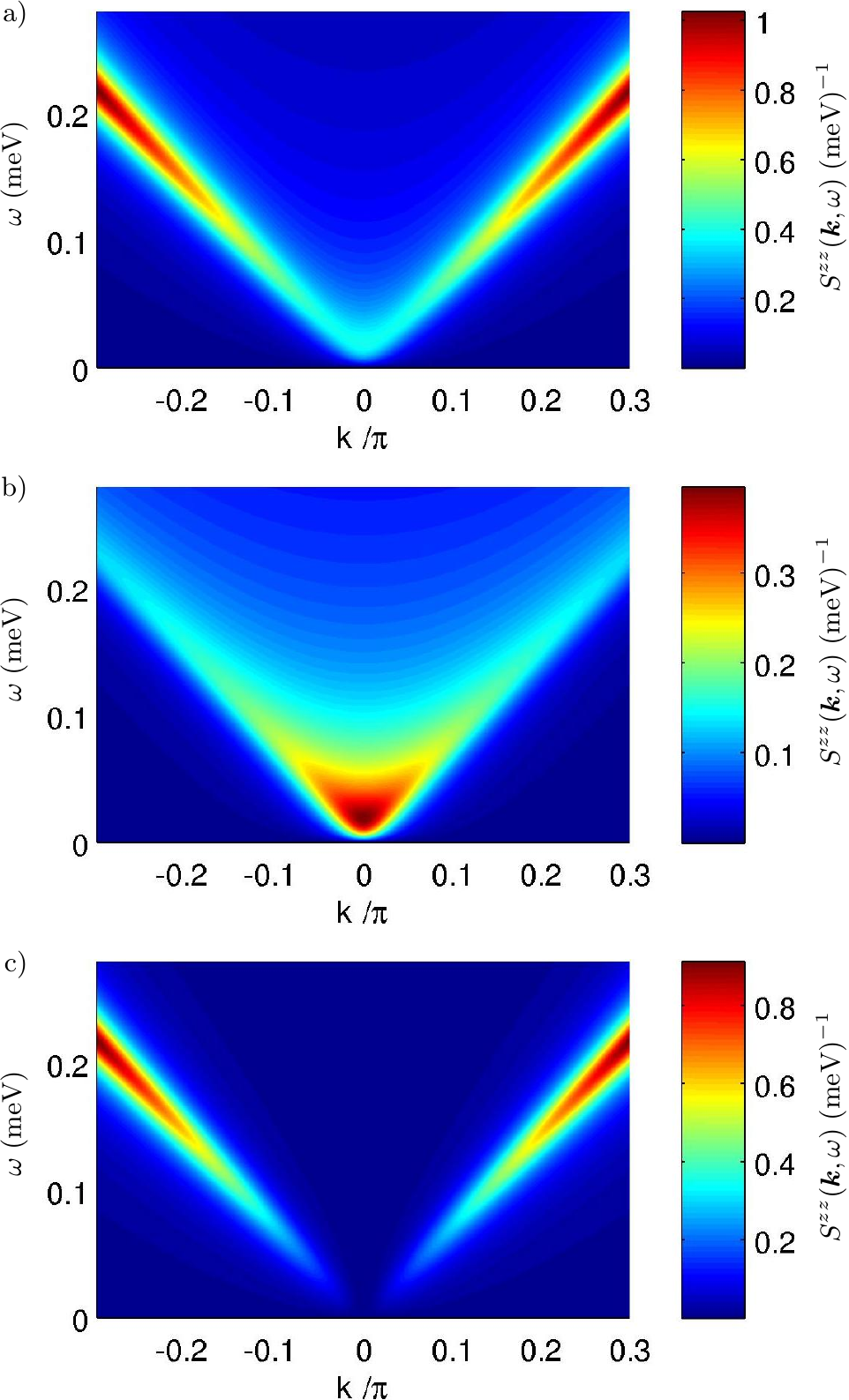}
\caption{(Color online)
Contour plots of the component $S^{zz}(\bd k,\omega)$ of the dynamic structure factor for $k_y =0$.
The upper panel (a) shows the total contribution that adds up from the anomalous contribution [middle panel, (b)] and the $\delta$-function contribution [lower panel, (c)].
The parameters and the smoothing procedure are the same 
as in Fig.~\ref{fig:szz}.
}
\label{fig:szzcontour}
\end{figure}

Because the weight of the $\delta$-function peaks 
in the transverse component of the structure factor 
in Eq.~(\ref{eq:Sxxyy})
is a factor of $( \Delta / \omega )^2$
larger than the $\delta$-function peak in the
longitudinal structure factor (\ref{eq:Szzt}), whereas the continuum
contributions have the same order of magnitude in both components, 
we conclude that the best way to detect
the anomalous scattering continua 
in Cs$_2$CuCl$_4$ 
is via a measurement of  the 
component   $S^{zz} ( \bd{k} , \omega ) $  of the
dynamic structure factor along the crystallographic $a$ axis
for magnetic fields below but not too close to $h_c$.
Note that for inelastic neutron scattering with unpolarized neutrons
the differential cross section always
involves a contribution from the transverse components of the 
structure factor,\cite{Marshall71}
 \begin{eqnarray}
 & & \frac{ d^2 \sigma ( \bd{k} , \omega ) }{ d \omega d \Omega} =  | f_{\bd{k}} |^2
 \sum_{ \alpha \beta} ( \delta_{\alpha \beta } - \hat{k}_{\alpha} \hat{k}_{\beta} )
 S^{\alpha \beta} ( \bd{k} , \omega ) 
 \nonumber
 \\
 &    = & | f_{\bd{k}} |^2
 \left[ ( 1 - \hat{k}_z^2 ) S^{zz} ( \bd{k} , \omega ) + ( 1 + \hat{k}_z^2 )
 S^{xx} ( \bd{k} , \omega ) \right],
 \hspace{7mm}
 \end{eqnarray}
where $ \hat{k}_\alpha = k_{\alpha} / | \bd{k} |$ and 
the magnetic form factor  $f_{\bd{k}}$ of the magnetic Cu$^{2+}$
ions in  Cs$_2$CuCl$_4$ has a rather weak momentum dependence.\cite{Coldea03}
The large $\delta$-function peaks of $ S^{xx} ( \bd{k} , \omega )$
will therefore dominate the neutron-scattering cross section, and it seems
rather difficult to detect the nonanalytic scattering continuum with unpolarized 
neutrons.  Note, however, that  longitudinal spin fluctuations in nickel
have been successfully detected with  polarized inelastic neutron scattering\cite{Boni91} so that
with polarized neutrons the longitudinal structure factor  of Cs$_2$CuCl$_4$
should be more easily accessible experimentally.
Contour plots of the total scattering intensities as well as
the anomalous and the $\delta$-function contributions
to the dynamic structure factor are shown in Fig.~\ref{fig:szzcontour}.
By comparing the upper panel (total intensity) with the lower panel
($\delta$-function contributions as also obtained in the framework of linear
spin-wave theory) the effect of the anomalous scattering continua can be seen.

\section{Summary and conclusions}
\label{sec:conclusions}

To summarize, we have shown that in the magnetically ordered ground state of
the anisotropic triangular lattice antiferromagnet Cs$_2$CuCl$_4$ in a uniform 
magnetic field along the crystallographic $a$ axis (cone state)
the
interactions between spin waves lead to infrareddivergent terms in the
$1/S$ expansion. These divergencies are generated by the coupling between
transverse and longitudinal fluctuations in the ordered phase.
Similar singularities
are expected to appear in any magnetically ordered spin system with
a linear magnon spectrum and a finite triangular vertex involving two
powers of the 
transverse spin-fluctuation field and one power
of the longitudinal spin-fluctuation field.
The reason why these divergencies have not been noticed in a previous
spin-wave calculation\cite{Veillette05} of the dynamic structure factor of
 Cs$_2$CuCl$_4$ is that in this calculation only the case of vanishing magnetic field
has been considered where the relevant triangular vertex vanishes.

We are not aware of any published neutron scattering data probing the
dynamic structure factor in the cone state of Cs$_2$CuCl$_4$ in an external
magnetic field.
We predict that in this case the spectral lineshape should
exhibit a characteristic anisotropy associated with the threshold divergence
proportional to 
$[ \omega^2 - c^2 \bd{k}^2 ]^{-1/2}$ of the anomalous contribution.
Although in the transverse components of the dynamic structure factor
the relative weight of this continuum is rather small,
it should be observable in the component 
 $S^{zz}(\bd k,\omega)$ associated with
spin correlations along the crystallographic $a$ axis.
Of course, non-singular higher-order terms in the $1/S$ expansion
also give rise to extended scattering continua,\cite{Veillette05}
which could overshadow
the continua due to the anomalous longitudinal fluctuations.
However, sufficiently close to the threshold $\omega = c | \bd{k}| $
the square-root divergence associated with the anomalous longitudinal fluctuations 
should be the dominant source of asymmetry of the spectral lineshape.

\section*{ACKNOWLEDGMENTS}
This work was financially supported by the DFG via SFB/TRR 49.
The work of PK was mostly 
carried out during a sabbatical stay
at the University of Florida, Gainesville; 
he would like to thank the University of Florida Physics Department
for its hospitality.
AK was supported by DOE Grant Nr. DE-FG02-05ER46236.

\end{document}